\documentstyle[11pt]{article}

\makeatletter
\def\numberbysection{\@addtoreset{equation}{section}
	\def\theequation{\thesection.\arabic{equation}}}
\makeatother

\makeatletter

\ifcase\@ptsize
 \font\tenmsy=msbm10
 \font\sevenmsy=msbm7
 \font\fivemsy=msbm5
\or
 \font\tenmsy=msbm10 scaled \magstephalf
 \font\sevenmsy=msbm8
 \font\fivemsy=msbm6
\or
 \font\tenmsy=msbm10 scaled \magstep1
 \font\sevenmsy=msbm8
 \font\fivemsy=msbm6
\fi

\newfam\msyfam
\textfont\msyfam=\tenmsy  \scriptfont\msyfam=\sevenmsy
  \scriptscriptfont\msyfam=\fivemsy

\def\hexnumber@#1{\ifnum#1<10 \number#1\else
 \ifnum#1=10 A\else\ifnum#1=11 B\else\ifnum#1=12 C\else
 \ifnum#1=13 D\else\ifnum#1=14 E\else\ifnum#1=15 F\fi\fi\fi\fi\fi\fi\fi}

\def\msy@{\hexnumber@\msyfam}

\mathchardef\lvertneqq="3\msy@00
\mathchardef\gvertneqq="3\msy@01
\mathchardef\nleq="3\msy@02
\mathchardef\ngeq="3\msy@03
\mathchardef\nless="3\msy@04
\mathchardef\ngtr="3\msy@05
\mathchardef\nprec="3\msy@06
\mathchardef\nsucc="3\msy@07
\mathchardef\lneqq="3\msy@08
\mathchardef\gneqq="3\msy@09
\mathchardef\nleqslant="3\msy@0A
\mathchardef\ngeqslant="3\msy@0B
\mathchardef\lneq="3\msy@0C
\mathchardef\gneq="3\msy@0D
\mathchardef\npreceq="3\msy@0E
\mathchardef\nsucceq="3\msy@0F
\mathchardef\precnsim="3\msy@10
\mathchardef\succnsim="3\msy@11
\mathchardef\lnsim="3\msy@12
\mathchardef\gnsim="3\msy@13
\mathchardef\nleqq="3\msy@14
\mathchardef\ngeqq="3\msy@15
\mathchardef\precneqq="3\msy@16
\mathchardef\succneqq="3\msy@17
\mathchardef\precnapprox="3\msy@18
\mathchardef\succnapprox="3\msy@19
\mathchardef\lnapprox="3\msy@1A
\mathchardef\gnapprox="3\msy@1B
\mathchardef\nsim="3\msy@1C
\mathchardef\napprox="3\msy@1D
\mathchardef\varsubsetneq="3\msy@20
\mathchardef\varsupsetneq="3\msy@21
\mathchardef\nsubseteqq="3\msy@22
\mathchardef\nsupseteqq="3\msy@23
\mathchardef\subsetneqq="3\msy@24
\mathchardef\supsetneqq="3\msy@25
\mathchardef\varsubsetneqq="3\msy@26
\mathchardef\varsupsetneqq="3\msy@27
\mathchardef\subsetneq="3\msy@28
\mathchardef\supsetneq="3\msy@29
\mathchardef\nsubseteq="3\msy@2A
\mathchardef\nsupseteq="3\msy@2B
\mathchardef\nparallel="3\msy@2C
\mathchardef\nmid="3\msy@2D
\mathchardef\nshortmid="3\msy@2E
\mathchardef\nshortparallel="3\msy@2F
\mathchardef\nvdash="3\msy@30
\mathchardef\nVdash="3\msy@31
\mathchardef\nvDash="3\msy@32
\mathchardef\nVDash="3\msy@33
\mathchardef\ntrianglerighteq="3\msy@34
\mathchardef\ntrianglelefteq="3\msy@35
\mathchardef\ntriangleleft="3\msy@36
\mathchardef\ntriangleright="3\msy@37
\mathchardef\nleftarrow="3\msy@38
\mathchardef\nrightarrow="3\msy@39
\mathchardef\nLeftarrow="3\msy@3A
\mathchardef\nRightarrow="3\msy@3B
\mathchardef\nLeftrightarrow="3\msy@3C
\mathchardef\nleftrightarrow="3\msy@3D
\mathchardef\divideontimes="2\msy@3E
\mathchardef\varnothing="0\msy@3F
\mathchardef\nexists="0\msy@40
\mathchardef\mho="0\msy@66
\mathchardef\thorn="0\msy@67
\mathchardef\beth="0\msy@69
\mathchardef\gimel="0\msy@6A
\mathchardef\daleth="0\msy@6B
\mathchardef\lessdot="3\msy@6C
\mathchardef\gtrdot="3\msy@6D
\mathchardef\ltimes="2\msy@6E
\mathchardef\rtimes="2\msy@6F
\mathchardef\shortmid="3\msy@70
\mathchardef\shortparallel="3\msy@71
\mathchardef\smallsetminus="2\msy@72
\mathchardef\thicksim="3\msy@73
\mathchardef\thickapprox="3\msy@74
\mathchardef\approxeq="3\msy@75
\mathchardef\succapprox="3\msy@76
\mathchardef\precapprox="3\msy@77
\mathchardef\curvearrowleft="3\msy@78
\mathchardef\curvearrowright="3\msy@79
\mathchardef\digamma="0\msy@7A
\mathchardef\varkappa="0\msy@7B
\mathchardef\hslash="0\msy@7D
\mathchardef\hbar="0\msy@7E
\mathchardef\backepsilon="3\msy@7F
\def\Bbb{\ifmmode\let\next\Bbb@\else
 \def\next{\errmessage{Use \string\Bbb\space only in math mode}}\fi\next}
\def\Bbb@#1{{\Bbb@@{#1}}}
\def\Bbb@@#1{\fam\msyfam#1}

\newfam\euffam
\font\sixeuf=eufm6
\font\eighteuf=eufm8
\font\twelveeuf=eufm10 scaled\magstep1
	\textfont\euffam=\twelveeuf
	\scriptfont\euffam=\eighteuf
  	\scriptscriptfont\euffam=\sixeuf
\def\euf{\fam\euffam\twelveeuf}

\makeatother

\newcommand{\cg}{{\euf g}}
\newcommand{\cgs}{{\euf g}^*}
\newcommand{\ZZ}{{\Bbb{Z}}}
\newcommand{\CC}{{\Bbb{C}}}
\newcommand{\RR}{{\Bbb{R}}}

\newcommand{\bge}{\begin{equation}}
\newcommand{\ee}{\end{equation}}
\newcommand{\sss}{\scriptscriptstyle}
\newcommand{\pdv}{\partial}
\newcommand{\cd}{{\cal D}}
\newcommand{\ch}{{\cal H}}
\newcommand{\wb}{W_b}
\newcommand{\wbc}{W_{b_0,c}}
\newcommand{\bc}{(b_0, ic)}
\newcommand{\hb}{{\cal H}_b}

\newcommand{\di}{{\rm Diff}S^1 / S^1}
\newcommand{\dis}{{\rm Diff}S^1 / SL^{(n)}(2,\RR)}
\newcommand{\dif}{{\rm Diff}S^1}
\newcommand{\dio}{{\rm Diff}_0 S^1}
\newcommand{\vvec}{\widehat{{\rm Vect}S^1}}
\newcommand{\vc}{{\rm Vect}S^1}

\newcommand{\lb}{{\cal L}_b}
\newcommand{\lbc}{{\cal L}_{b_0,c}}
\newcommand{\vir}{\widehat{{\rm Diff}S^1}}
\newcommand{\kl}{K\"{a}hler }
\newcommand{\aads}{{\rm Ad}^*}
\newcommand{\ad}{{\rm ad}}
\newcommand{\ads}{{\rm ad}^*}
\newcommand{\aadsg}{{\rm Ad}^*_g}
\newcommand{\aad}{{\rm Ad}}
\newcommand{\aadg}{{\rm Ad}_g}
\newcommand{\tu}{\tilde{u}}
\newcommand{\tv}{\tilde{v}}
\newcommand{\hu}{\hat{u}}
\newcommand{\hv}{\hat{v}}
\newcommand{\hj}{\hat{J}}

\newcommand{\hln}{\hat{L}_n}
\newcommand{\hl}{\hat{L}}
\newcommand{\pu}{\Phi_u}
\newcommand{\pn}{\Phi_n}
\newcommand{\xu}{\xi_u}
\newcommand{\xn}{\xi_n}
\newcommand{\bb}{{\bf b}}
\newcommand{\bbo}{{\bf b_0}}
\newcommand{\bo}{b_0}
\newcommand{\eps}{\epsilon}
\newcommand{\azz}{\alpha(z, \bar{z})}
\newcommand{\bzz}{\beta(z, \bar{z})}
\newcommand{\azzp}{\alpha(z', \bar{z}')}
\newcommand{\bzzp}{\beta(z', \bar{z}')}
\newcommand{\mzz}{\mu_n(z, \bar{z})}
\newcommand{\rzz}{\rho(z, \bar{z})}
\newcommand{\gzz}{\gamma(z, \bar{z})}
\newcommand{\co}{{\cal O}}
\newcommand{\nul}{|\:\rangle}
\newcommand{\nuls}{|s\rangle}
\newtheorem{prop}{Proposition}

\begin{document}
\begin{titlepage}
\begin{center}
April 21, 1992     \hfill    LBL-33210 \\
          \hfill    UCB-PTH-92/09 \\

\vskip .5in

{\large \bf Virasoro Representations on $\di$ Coadjoint Orbits}
\footnote{This work was supported by the Director, Office of Energy
Research, Office of High Energy and Nuclear Physics, Division of High
Energy Physics of the U.S. Department of Energy under Contract
DE-AC03-76SF00098.}

\vskip .5in

Washington Taylor IV\footnote{\tt wati@physics.berkeley.edu}\\[.5in]

{\em  Department of Physics\\
      University of California\\
      and\\
      Theoretical Physics Group\\
      Lawrence Berkeley Laboratory\\
      1 Cyclotron Road\\
      Berkeley, California 94720}
\end{center}

\vskip .5in

\begin{abstract}
A new set of realizations of the Virasoro algebra on a bosonic Fock
space are found by explicitly computing the Virasoro representations
associated with coadjoint orbits of the form $\di$.  Some progress is
made in understanding the unitary structure of these
representations.  The characters of these representations are exactly
the bosonic partition functions calculated previously by Witten using
perturbative and fixed-point methods.  The representations
corresponding to the discrete series of unitary Virasoro
representations with $c \leq 1$ are found to be reducible in this
formulation, confirming a conjecture by Aldaya and Navarro-Salas.
\end{abstract}
\end{titlepage}
\renewcommand{\thepage}{\roman{page}}
\setcounter{page}{2}
\mbox{ }

\vskip 1in

\begin{center}
{\bf Disclaimer}
\end{center}

\vskip .2in

\begin{scriptsize}
\begin{quotation}
This document was prepared as an account of work sponsored by the United
States Government.  Neither the United States Government nor any agency
thereof, nor The Regents of the University of California, nor any of their
employees, makes any warranty, express or implied, or assumes any legal
liability or responsibility for the accuracy, completeness, or usefulness
of any information, apparatus, product, or process disclosed, or represents
that its use would not infringe privately owned rights.  Reference herein
to any specific commercial products process, or service by its trade name,
trademark, manufacturer, or otherwise, does not necessarily constitute or
imply its endorsement, recommendation, or favoring by the United States
Government or any agency thereof, or The Regents of the University of
California.  The views and opinions of authors expressed herein do not
necessarily state or reflect those of the United States Government or any
agency thereof of The Regents of the University of California and shall
not be used for advertising or product endorsement purposes.
\end{quotation}
\end{scriptsize}

\vskip 2in

\begin{center}
\begin{small}
{\it Lawrence Berkeley Laboratory is an equal opportunity employer.}
\end{small}
\end{center}

\newpage
\renewcommand{\thepage}{\arabic{page}}
\setcounter{page}{1}

\numberbysection
\section{Introduction}
\baselineskip 18.5pt

The method of coadjoint orbits originated by Kirillov and Kostant
twenty years ago \cite{Kir} has proven to be a valuable tool in
investigating geometrical aspects of the representation theory of Lie
groups.  The Kirillov-Kostant approach
is essentially a generalization of the Borel-Weil theorem, which
constructs irreducible unitary representations of a finite-dimensional
compact semi-simple Lie group $G$ as spaces of holomorphic sections of
complex line bundles over the homogeneous space $G/T$, where $T$ is a
maximal subtorus of $G$.  In the coadjoint orbit approach, one begins
with a group $G$, with Lie algebra ${\cg}$.  The group $G$ has a
natural coadjoint action on the dual space $\cg^*$.  Choosing an
element $b$ in $\cg^*$, one considers the coadjoint orbit $W_b$ of $b$
in $\cg^*$.  For any $b$, the space $W_b$ has a natural symplectic
form $\omega$.  For those $b$ with the property that a complex line
bundle ${\cal L}_b$ can be constructed over $W_b$
with curvature form $i \omega$, one attempts to relate an appropriate
space of sections of
${\cal L}_b$ to an irreducible
unitary representation of $G$ by using the technique of geometric
quantization on the space $W_b$.  For finite-dimensional compact
semi-simple $G$, the representations produced by
this construction are equivalent to those given by the Borel-Weil
theory.
The coadjoint orbit approach is particularly useful in the case of
non-compact groups, where the Borel-Weil theory does not apply.
It is possible to apply the Borel-Weil approach to certain infinite-dimensional
groups such as the centrally extended loop groups
$\widehat{LG}$ \cite{PS}.  For other infinite-dimensional groups, such
as the (orientation-preserving) diffeomorphism group of the circle
${\rm Diff} S^1$, and its central extension $\vir$, the Virasoro
group, there are difficulties with applying even the more general
coadjoint orbit theory.
Many of the Virasoro coadjoint orbits do not admit a \kl
structure, so that it is difficult to geometrically quantize these
spaces.  Also, it is known that the Virasoro group has rather peculiar
mathematical properties, such as the fact that the exponential map on
the Lie algebra is neither onto nor 1-1 in the vicinity of the
identity.  Due to these difficulties, a full understanding of the
coadjoint orbit representations for this group has not yet been
attained, although there are some partial results in this
direction \cite{lp,segal,Witt1}.  Achieving an understanding of the
geometry of the coadjoint orbit representations of the Virasoro group
could be a valuable step in the general study of Virasoro
representations and conformal field theory.  In particular, recent
work  \cite{ans2,al-sh1,ber-og} indicates the existence of a
relationship between these
Virasoro representations and the $SL(2,\RR)$ current algebra found by
Polyakov in 2-d gravity \cite{Poly}.

The coadjoint orbits of $\vir$ were classified by Segal \cite{segal}
and Lazutkin and Pankrotova \cite{lp}.  The Lie algebra of $\dif$ is
the space $\vc$ of smooth vector fields on $S^1$, and the natural dual
to this space
is the space of smooth quadratic differentials on $S^1$.  The coadjoint
orbits of $\vir$ can be obtained by finding the
stabilizers in $\vir$ of a general quadratic differential $b$ on
$S^1$.  Among other spaces, one finds that $\di$ and $\dis$ can appear
as coadjoint orbit spaces of $\vir$, where $SL^{(n)}(2,\RR)$ is
the subgroup of $\dif$ generated by
the Virasoro generators $L_0, L_{\pm n}$.  Witten \cite{Witt1} has made
some progress in relating these coadjoint orbits to the irreducible
unitary representations of $\vir$, which have previously been classified
using algebraic methods \cite{FF,FQS,GKO}.  By using perturbative
techniques and the fixed point version of the Atiyah-Singer index
theorem, Witten was able to calculate the characters of the
representations associated with the $\di$ orbits, which he found to be
the standard bosonic partition function associated with a Virasoro
representation with no null states.  The perturbative methods used by
Witten, however, are only valid in the semi-classical $c \gg 1$
domain.  In particular, the structure of the $c \leq 1$ discrete
series of unitary representations could not be understood in terms of
coadjoint orbits using these techniques.  Witten conjectured that
these representations would be found in the $\dis$ orbits, but since
these spaces do not admit \kl structures, it has not yet been possible
to perform geometric quantization in these cases, and the
representations associated with these orbits are still not understood.

More recently, related investigations have provided clues to the structure of
the Virasoro coadjoint orbit representations.  By using a technique
involving quantization on a group manifold, Aldaya and Navarro-Salas
were able to construct representations of the Virasoro group on spaces
of polarized functions on the group manifold $\vir$
itself \cite{ANS1}.  For those values of $c$ and $h$ where the Kac
determinant vanishes (i.e., where the algebraically constructed
representation contains a null state), they made the interesting
observation that the representation constructed through their method
is reducible, yet
contains only a single highest weight vector.  By taking the orbit of
the highest weight vector under the Virasoro action, they get a
subspace of the original representation space which corresponds
exactly to the appropriate irreducible unitary representation in the
$c \leq 1$ discrete series;  they did not, however, investigate the
existence of unitary structures on their representation spaces.  By
analogizing their techniques to the coadjoint orbit method, Aldaya and
Navarro-Salas conjectured that a similar situation would arise in the
 $\di$ coadjoint orbit representations with $c \leq 1$.  In this paper we show
explicitly that this is indeed the case.

Another interesting and related approach was taken by Alekseev and
Shatashvili \cite{al-sh1}.  They constructed a natural set of quantum
field theories, parameterized by $h$ and $c$, corresponding to quantum
mechanical systems on the group manifold $\dif$.  These quantum field
theories are all symmetric under $S^1$, and thus can be viewed as
theories on the coadjoint orbit space $\di$.  These field theories are
constructed in such a fashion that their Hilbert spaces should
naturally be associated with the modules carrying coadjoint orbit
Virasoro representations.  When $h = \frac{-c(n^2-1)}{24}$, with $n$ a
non-zero integer, these field theories are symmetric under
$SL^{(n)}(2,\RR)$, and thus contain a residual symmetry when viewed as
theories on $\di$.  By changing the domain of the fields from $S^1$ to
$\RR$, effectively dropping the periodicity requirement, Alekseev and
Shatashvili came up with a closely related set of theories, all of
which have the extra $SL(2,\RR)$ symmetry.  In the case where $h =
c/24$, Alekseev and Shatashvili showed that the action of their field
theory coincides with the gravitational Wess-Zumino-Witten (WZW)
action, and they used this relationship to interpret the $SL(2,\RR)$
symmetry of the associated 2d gravity theory in a natural way.  In a
later paper \cite{al-sh2}, Alekseev and Shatashvili investigated the
structure of Virasoro representations by yet another geometrical
method involving quantization of the ``model space'' of the Virasoro
group.  Finally, Aldaya, Navarro-Salas, and Navarro have also
considered a field theory model like that developed by Alekseev and
Shatashvili, from their approach of quantization on the group manifold
\cite{ans2}.  They achieve a natural understanding of the hidden
$SL(2,\RR)$ symmetry in the gravitational WZW model in terms of the
separate left- and right-invariant vector fields on the group
manifold.

The goal of this paper is to explicitly construct the Virasoro
representations associated with the $\di$ coadjoint orbits.  For every
$c$ and $h$ such that $\frac{c - 24h}{c}$ is not the square of a positive
integer, a representation on such an orbit exists.  For the
exceptional values of $c$ and $h$, our formulae still give
representations, but in these cases the representations cannot be
directly interpreted as arising from coadjoint orbits.
The representations are constructed by putting a countable set of
holomorphic coordinates
$z_1, z_2, \ldots$ on $\di$, and explicitly computing the action
of the Virasoro generators $\hln$ on the space $R$ of polynomials in
the $z_i$'s ($R$ is the space of holomorphic sections of an
appropriate line bundle over $\di$.)  The explicit calculation of the
operators $\hln$ is accomplished by making a judicious choice of
gauge, in which a connection for the desired line bundle can be
calculated through a simple recursive procedure.  In these
representations, the generators $\hln$ act as first-order differential
operators on the space $R$.  Although this is a necessary consequence
of the general form of the coadjoint orbit construction, this is in
some sense a surprising result; most standard representations of the
Virasoro algebra in terms of free fields, such as those developed by
Feigen and Fuchs \cite{FF2}, involve second order derivatives in some
of the generators, when the free fields are rewritten in terms of
variables and derivatives.  Although the generators $\hln$ are
expressed as formal power series with an infinite number of terms, the
action of any generator on a fixed polynomial in $R$ only involves a
finite number of terms, and is computable.

Once we have
constructed the $\di$ representations explicitly, certain aspects of their
structure become quite apparent.  For all these representations, the
character of the representation is easily seen to be exactly the
bosonic partition function calculated perturbatively by Witten, since
all polynomials in the $z_i$'s appear in the representation space.
For $c$ and $h$ corresponding to the discrete unitary series, one
also finds that the representations have exactly the structure predicted
by Aldaya and Navarro-Salas.  The relationship of these explicit forms
for the $\di$ representations to the field theory approach used
in \cite{ans2,al-sh1} is not yet understood.  It seems, however, that
it should be possible to describe the field theory of Alekseev and
Shatashvili in terms of the representations described here, with the
ring $R$ becoming the Hilbert space of the quantum theory.
This
approach could lead to a purely algebraic construction of the theory
of 2d gravity.  Work in this direction is currently in progress.

The structure of the rest of this paper is as follows:  In Section 2,
we review the coadjoint orbit approach to constructing
representations, and prove several propositions which will justify the
``gauge-fixing'' procedure we use to construct explicit
representations.  As an example, we apply this procedure to the group
$SU(2)$.  In section 3, we carry out the construction in the case of
the $\di$ Virasoro orbits, and we discuss the question of unitarity
for the resulting representations.  Many of the calculations in this
section are carried out in a rather formal fashion, without regard to
convergence issues and other technicalities related to the
infinite-dimensional nature of the Virasoro group.  It is presumed
that the analysis involved could be reformulated in a more rigorous
mathematical language, however we have not attempted to do so here
beyond a few brief and necessary digressions.  In section 4, we review
the salient
features of the representations we have constructed, and relate
the results of this paper to other recent work.

\section{Coadjoint Orbits and Representations}
\baselineskip 18.5pt

In this section we review the coadjoint orbit approach to group
representations, and prove several results which will be essential to
our construction of Virasoro representations in section 3.
In the introductory paragraphs of this section several standard
results on coadjoint orbits are stated without proof; the
verifications of these statements are fairly
straightforward algebraic manipulations.  Otherwise, an attempt has
been made to
make this paper relatively self-contained.  For a more comprehensive
introduction to the coadjoint orbit approach to representation theory,
the reader should consult Kirillov \cite{Kir} or
Witten \cite{Witt1}.

Given
a group $G$ with lie algebra $\cg$, consider the space $\cg^*$ dual
to $\cg$.  The adjoint action of $G$ on $\cg$ associates with each $g
\in G$ a map
\bge {\rm Ad}_g:\cg \rightarrow \cg.\ee
When $G$ is a matrix group, one has
\bge {\rm Ad}_g:u \mapsto g u g^{-1}.\ee
$G$ also has a dual action on $\cg^*$, denoted ${\rm Ad}^*$, where
\bge \langle {\rm Ad}_g^* b, u \rangle = \langle b, {\rm Ad}_{g^{-1}}
u \rangle, \;\;\;{\rm for}\; b \in \cg^*, u \in \cg.
\ee
The action ${\rm Ad}^*$ is referred to as the coadjoint action of $G$
on $\cg^*$.  The derivative of the adjoint action gives an action of
$\cg$ on $\cg$, denoted by $\ad$, where $\ad_u v = [u,v]$, for all $u,
v \in \cg$.  Similarly, the infinitesimal coadjoint action of $\cg$ on
$\cgs$ is
denoted $\ads$, and is given by
\bge
\langle \ads_v b, u\rangle  = \langle b, [u,v] \rangle , \;\;\;{\rm
for}\; b \in \cg^*, u,v \in \cg.\ee
For any $b \in \cg^*$, one can consider its orbit $W_b$ in
$\cgs$ under the coadjoint action of $G$.  It turns out that $W_b$ admits
a natural symplectic structure, which may be defined as follows:
There is a natural association between
elements of $\cg$ and tangent vectors to $\wb$ at $b$.  Given an
element $u \in \cg$, we define $\tu (b) \in T_b \wb$ to be the tangent
vector to $\wb$ at $b$ associated with $\ads_u b$.  (Note that $\tu(b)
= 0$ when $u$ is in the stabilizer of $b$; i.e., when $\ads_u b = 0$.)
We can define a 2-form $\omega$ on $\wb$ by
\bge\omega(\tu(b),\tilde{v}(b)) = \langle b, [u,v] \rangle.
\label{eq:omega} \ee
It can be verified that this 2-form is well-defined, closed,
$G$-invariant, and nondegenerate, and thus defines a $G$-invariant
symplectic structure on $W_b$.  $\omega$ also gives a Poisson bracket
structure to the space of functions on $\wb$.  In component notation,
the Poisson
bracket of two functions $f$ and $g$ is given by
\bge
\lbrace f,g \rbrace = \omega^{ij} (\pdv_i f)(\pdv_j g),\ee
where $\omega^{ij}$ are the components of $\omega^{-1}$.  Every
function $f$ on $\wb$ generates a Hamiltonian vector field $v_f$ on
$\wb$, defined by
\bge v_f^i = \omega^{ij} (\pdv_j f).\ee
For any $u \in \cg$, there is a function $\pu$ on $W_b$ which generates
the Hamiltonian vector field $\tu$.  This function is given by
\bge
\pu(b) = - \langle b, u \rangle .\ee
To see that $\pu$ generates the vector field $\tu$, we use the fact
that for any $v \in \cg$,
\bge
\tilde{v}^j\pdv_j \pu(b) = - \langle b, [u,v] \rangle = \omega_{jk}
\tilde{v}^j \tu^k.\ee
Since the vector fields $\tv$ span the tangent space to $\wb$ at $b$,
we have
\bge
\pdv_j \pu(b) = \omega_{jk} \tu^k (b),\ee
so
\bge  \omega^{ij}\pdv_j \pu(b) = \tu^i.\ee
The functions $\pu$ also satisfy the equation
\bge
\lbrace \pu, \Phi_v \rbrace = \Phi_{[u,v]},
\label{eq:poiss} \ee
since
\begin{eqnarray}
\lbrace \pu, \Phi_v \rbrace & = & \omega^{ij} (\pdv_i \pu) (\pdv_j
\Phi_v)
= \omega^{ij} (\omega_{ik} \tu^k)( \omega_{jl} \tilde{v}^l) \nonumber \\
& = & \omega_{ik} \tu^k \tilde{v}^i
 =  \langle b, [v,u] \rangle
 =  \Phi_{[u,v]}.
\end{eqnarray}
In order to construct representations of $G$ using the coadjoint orbit
$\wb$, it is now necessary to quantize the manifold $\wb$ according to
the technique of geometric quantization \cite{Wood,Sni}.  The first
step in this procedure is to construct a complex line bundle $\lb$
over $\wb$ with curvature form $i \omega$.  This is known as
``prequantization''. For this step to be possible, it
is necessary that $\frac{\omega}{2\pi}$ be an integral cohomology
class (i.e., that the integral of $\omega$ over any closed 2-surface
in $\wb$ be an integral multiple of $2\pi$.)   If such a line bundle
$\lb$ exists,
then there is a natural homomorphism $\phi$ from the Lie algebra $\cg$
to the space of first-order differential operators on sections of
$\lb$, given by
\bge
\phi : u \mapsto \hu = -\nabla_{\tu} +i \pu,\ee
where $\nabla_{\tu}$ is the covariant derivative in $\lb$ in the
direction $\tu$.  Explicitly, written in component notation in a local
coordinate chart,
\bge
\hu = -\tu^i(b)(\pdv_i + A_i(b)) +i \pu(b),
\label{eq:oprep}
\ee
where $A_i$ is a connection on $\lb$ satisfying $\pdv_i
A_j - \pdv_j A_i = i \omega_{ij}$.  To verify that $\phi$ is a
homomorphism, we must
check that
\bge
\lbrack \hu, \hv \rbrack = \widehat{\lbrack u,v \rbrack}.
\label{eq:comm}
\ee
We define $\xi_u$ to be the differential operator corresponding to the
vector field $-\tu$; i.e., $\xi_u = -\tu^i \pdv_i$, and we define
$A_u = \tu^i A_i$.  With these definitions,
\bge
\hu = \xu - A_u +i \pu.\label{eq:oprep2} \ee
One can easily calculate
\bge
\lbrack \xu, \xi_v \rbrack = \xi_{[u,v]},\ee
and
\bge  \xu \Phi_v(b) = \Phi_{[u,v]}(b).\ee
One also finds that
\begin{eqnarray}
\xu A_v - \xi_v A_u & = & \widetilde{[u,v]}^i A_i - \tu^i \tilde{v}^j
(\pdv_i A_j - \pdv_j A_i) \nonumber \\
& = & A_{[u,v]} + i \Phi_{[u,v]}. \label{eq:xacomm} \end{eqnarray}
Note that since the vectors $\tu$ span the tangent space to $\wb$ at
each point, Equation \ref{eq:xacomm}, along with the conditions that
$A_u$ is linear in $u$ and that $A_u(b) = 0$ when $\ads_u b = 0$,
could have been taken as the definition of a connection $A_u$
associated with the derivative operators $\xu$.  It is now trivial to
compute the commutator
\begin{eqnarray}
[\hu, \hv] & = & [\xu - A_u +i \pu, \xi_v - A_v +i \Phi_v] \nonumber \\
& = & \xi_{[u,v]} - A_{[u,v]} +i \Phi_{[u,v]} \nonumber \\
& = & \widehat{[u,v]}.  \end{eqnarray}
Thus $\phi$ is a
homomorphism, so we have determined that $\phi$ gives a representation
of $\cg$ on the space of smooth sections of $\lb$.  Unfortunately,
this representation is in general much too large to be irreducible;
this is where the second stage of geometric quantization enters,
which involves choosing a ``polarization''.  We will only be concerned
here with a specific type of polarization, the \kl polarization.  In
general, choosing a polarization restricts the space of
allowed smooth sections of $\lb$ to a subspace containing only those
sections which satisfy some local first-order differential equations.
A \kl polarization of $\wb$ exists when $\wb$ admits a $G$-invariant
\kl structure with $- \omega$ as the associated $(1,1)$-form.  This
condition is equivalent to the condition that $\wb$ admits a
$G$-invariant complex structure with respect to which $\omega$ is a
$(1,1)$-form; i.e., the only nonvanishing terms in $\omega$ have one
holomorphic and one antiholomorphic index.  Note that $-\omega$ is
usually constrained to be a positive form as part of the \kl
condition; since this property is not necessary for the construction
of representations, we will not impose it here.  In general, if $\wb$ does
not admit a \kl polarization, and is not equivalent to a cotangent
bundle, there is no standard way to find a polarization, and carrying
out the geometric quantization program becomes extremely difficult.
In case $\wb$ does admit a \kl polarization, we can restrict the space
of allowed sections of $\lb$ to the space $\hb$ of holomorphic sections.
When $\lb$ has a Hermitian metric, then we can restrict $\hb$ to be the
Hilbert space of square-integrable holomorphic sections of $\lb$.
According to the general principles of Kirillov and Kostant, the
action of $G$ on $\hb$ should give an irreducible unitary
representation of $G$ for every $b$ such that $\hb$ can be
constructed.  This principle holds fairly well for compact
semi-simple finite-dimensional groups, and even for loop groups,
however it does not seem to hold in complete generality.  Some of the
representations
of $\vir$ constructed this fashion are nonunitary, and some are
reducible.

Now that we have reviewed the standard approach to constructing
representations via coadjoint orbits, we can prove several assertions
which will simplify the process of explicitly constructing these
representations in local coordinates.  If one attempts to use Equation
\ref{eq:oprep2} to construct explicit formulae for the operators $\hu$
as differential operators on $\hb$, one encounters several obstacles.
First, it is necessary to calculate the functions $\pu$ in local
coordinates.  Second, one must find an explicit formula for
a connection $A_u$ which satisfies (\ref{eq:xacomm}).  Finding these
expressions in terms of a local set of holomorphic coordinates is in
general a somewhat nontrivial problem.  Note, however that the
operator $\hu$ can be written as
\bge
\hu = \xu + f_u,\ee
where $\xu$ is the first-order differential operator defined above,
and $f_u$ is a function of the local coordinates satisfying
\bge
\xu f_v - \xi_v f_u = f_{[u,v]}.\label{eq:xfcomm} \ee
We will find it easiest to construct explicit expressions for the
operators $\hu$ by finding directly a set of functions $f_u$ which
satisfy (\ref{eq:xfcomm}), and which correspond to the representation in
question.  We find these functions $f_u$ by making a simplifying
assumption which amounts to choosing a simple gauge for the connection
$A_u$.  To ensure that the set of $f_u$'s we construct in this fashion
are equivalent to those we would get from (\ref{eq:xacomm}) by a
specific choice of gauge, we will need the following two propositions.

\begin{prop}
Given a coadjoint orbit $\wb$ of a group $G$, with $\lb$ a complex
line bundle
over $\wb$ with curvature $i \omega$, and with $\xu$ and $\pu$ defined
as above, on a coordinate chart corresponding to a
local trivialization of $\lb$, if a set of functions $f_u$ on $\wb$
are linear in $u \in \cg$, and satisfy the conditions
\vspace{.1in} \newline
\makebox[.3in][r]{{\rm (}{\sl i}\/{\rm )}} \hspace{.08in}
\parbox[t]{5.6in}
{$\xu f_v - \xi_v f_u = f_{[u,v]}$,}
\vspace{.1in} \newline
\makebox[.3in][r]{{\rm (}{\sl ii}\/{\rm )}} \hspace{.08in}
\parbox[t]{5.6in}
{$f_u(b) = i \pu(b)$ when $\ads_u b = 0$,}
\vspace{.1in} \newline
then the operators $\hu = \xu + f_u$ are equal to the operators $\hu$
{}from Equation \ref{eq:oprep2} for some choice of connection $A_u$ on
$\lb$ satisfying {\rm (}\ref{eq:xacomm}{\rm )}.
\label{p:p1}
\end{prop}
\noindent {\it Proof.} \,\,
To prove this proposition, it will suffice to show that the functions
$A'_u(b) = - f_u(b) + i \Phi_u(b)$ satisfy (\ref{eq:xacomm}), are linear
in $u$, and are zero when $\ads_u b = 0$.  The last two conditions
follow immediately from the definition of $f_u$ and assumption ({\sl
ii}).  To see that $A'_u$ satisfies (\ref{eq:xacomm}) is a simple
calculation:
\begin{eqnarray}
\xu A'_v - \xi_v A'_u & = & \xu (- f_v + i\Phi_v) - \xi_v (- f_u + i \pu)
\nonumber \\
& = & - f_{[u,v]} + 2 i \Phi_{[u,v]} \nonumber \\
& = & A'_{[u,v]} + i \Phi_{[u,v]}. \end{eqnarray}
Thus, $A'_u$ is a valid connection on $\lb$, and the proposition is proven.
$\Box$

\begin{prop}
With the same premises as Proposition \ref{p:p1}, when $G$ is path
connected the condition {\rm (ii)} can be replaced by the weaker condition
\vspace{.1in} \newline
\makebox[.3in][r]{{\rm (}{\sl ii$'$}\/{\rm )}} \hspace{.08in} \parbox[t]{5.6in}
{For some point $b_0 \in \wb$, $f_u(b_0) = i \pu(b_0)$ for all $u$
such that $\ads_u b_0 = 0$,}
\vspace{.1in} \newline
and the result of proposition \ref{p:p1} still holds.
\label{p:p2}
\end{prop}
\noindent {\it Proof.} \,\,
We need to prove that when
$G$ is path connected, condition ({\sl ii$'$}) implies condition ({\sl
ii}).  Assume $\ads_u b = 0$ for some $u \in \cg, b \in \wb$.  Since
$b_0 \in \wb$, for some $g \in G$ we have $b = \aadsg b_0$.  If $u$
stabilizes $b$, then $u_0 = \aad_{g^{-1}} u$ must stabilize $b_0$.
But then we have
\bge
\langle b, u\rangle = \langle \aadsg b_0 ,\aadg u_0\rangle = \langle
b_0 , u_0\rangle,\ee
so $\pu(b) = \Phi_{u_0}(b_0)$.  It remains to be shown that $f_u(b) =
f_{u_0}(b_0)$.  Since $G$ is path connected, we have a path $g(t)$ in
$G$ with $g(0) = 1$ and $g(1) = g$.  We
claim that
\bge
\frac{d}{dt} f_{u(t)}(b(t)) = 0,\ee
where $u(t) = \aad_{g(t)} u_0$, and $b(t) = \aads_{g(t)} b_0$.
Defining
\bge v(t) = \frac{d g(t)}{d t} g^{-1}(t) \in \cg,\ee
we have
\bge
\frac{d}{dt} b(t) = \ads_v b(t),\ee
and
\bge \frac{d}{dt} u(t) = \ad_v u(t).\ee
It follows that
\begin{eqnarray}
\frac{d}{dt} f_{u(t)}(b(t)) & = & -\xi_v f_{u(t)} (b(t)) +
f_{[v,u(t)]}(b(t)) \nonumber \\
& = & -\xi_v f_{u(t)} (b(t)) + \xi_{u(t)} f_{v} (b(t)) +
f_{[v,u(t)]}(b(t)) \nonumber \\
& = & 0,  \end{eqnarray}
where we have used the fact that $\tu(t)(b(t)) = 0$.  Thus, we have
shown that
\bge f_u(b) = f_{u_0}(b_0) = i \Phi_{u_0}(b_0) = i \pu(b).\ee
Since $u$ and $b$ were an arbitrary solution of $\ads_u b = 0$, we
have proven that condition ({\sl ii$'$}) implies condition ({\sl
ii}), and thus the proposition is proven.$\Box$

We will now as an example use the coadjoint orbit approach to
construct representations of $SU(2)$.  For a similar discussion from
the point of view of the Borel-Weil theorem, see Alvarez, Singer, and Windey
\cite{Alv1}.  Take the generators of the algebra $\cg = su(2)$ to be
$\{iJ_k: k=1,2,3\}$, where $[J_j, J_k] = i \eps_{jkl} J_l$.  $\cg$ is
a three-dimensional real vector space.  Taking coordinates $x^1, x^2,
x^3$ on $\cg$, an arbitrary element $u \in \cg$ can be written as $u =
i \Sigma x^k J_k$.  An arbitrary element $g$ of $G$ can be written as
$g = e^{u}$, where $u \in \cg$.  In a vicinity of the identity, this
description of $g$ is unique.  The adjoint representation of $G$ acts
on $\cg$ via rotations which preserve the Euclidean scalar product;
the generator $iJ_k$ corresponds to rotation about the $x^k$ axis.
Since $\cg$ is finite-dimensional, $\cgs$ can be identified with $\cg$
using the Euclidean scalar product.  Under this identification, the
coadjoint action of $G$ on $\cgs$ is also given by rotations.  Given a
vector ${\bf b} = (b_1, b_2, b_3) \in \cgs$, where $\langle \bb, iJ_k
\rangle$ is defined to be $b_k$, the coadjoint orbit of $\bb$ is given
by
\bge
\wb = \{ \bb' \in \cgs: |\bb'|^2 = b^2\},\ee
which is just the 2-sphere in $\cgs$ of radius $b = |\bb|$.  We will
now explicitly calculate the 2-form $\omega$ on $\wb$.  We choose a
canonical element $\bbo = (0,0,b) \in \wb$.  To calculate $\omega$ at
the point $\bbo$, we need only find the explicit correspondence
between elements of $\cg$ and $T_{\bbo}W_b$.  Under the Lie algebra
coadjoint action, we have
\begin{eqnarray}
\ads_{iJ_1} \bbo & = & (0,b, 0), \nonumber \\
\ads_{iJ_2} \bbo & = & (-b,0, 0),  \\
\ads_{iJ_3} \bbo & = & (0,0, 0). \nonumber \end{eqnarray}
It follows that
\bge  \omega_{12}(\bbo) = \langle \bbo,  [- \frac{iJ_2}{b},
\frac{iJ_1}{b}] \rangle =
-\frac{1}{b}.\ee
Since $\omega$ is $G$-invariant, it
is easy to see that $\omega$ is defined globally on $\wb$ by
\bge \omega_{ij}(\bb) = - \frac{1}{b^2} \eps_{ijk} b_k.\ee
In order for $\omega/2\pi$ to be an integral form, we must have
$\int_{\wb} \omega / 2 \pi = - 2b \in \ZZ$, so $b$ must be a half-integer.
Thus, whenever $b \in \ZZ/2$, we can construct a line bundle $\lb$
over $\wb$ with curvature form $i \omega$.

We would now like to find a
$G$-invariant \kl structure on $\wb$ compatible with $\omega$, so that
we can restrict attention to holomorphic sections of $\lb$, according
to the prescription of geometric quantization.  A standard result from
group theory allows us to describe this complex structure from an
algebraic viewpoint which will be useful in the case of the Virasoro
group.  For a similar approach to this construction see Zumino
\cite{Zum}.

Consider the stabilizer $H$ in $G$ of $\bbo$ under the coadjoint
$G$-action on $\cgs$.  $H$ is clearly just the $U(1)$ subgroup
generated by $iJ_3$,
\bge  H = \{ e^{itJ_3}: t \in \RR\}.\ee
There is a 1-1 correspondence between points in $\wb$ and $G/H$, since
for every $\bb \in \wb$, there exists a $g \in G$ such that $\bb =
\aadsg \bbo$, and for $g, g' \in G$,
\bge
\aadsg \bbo = \aads_{g'} \bbo \;\; {\rm iff} \;\; g = g'h \;\; {\rm
for}\;{\rm some}
 \;\; h \in H.\ee
Note that the coadjoint action of $G$ on $\wb$ corresponds to the left
action of $G$ on $G/H$.  We
will now describe a natural complex structure on $G/H$.  If we define
\bge
J_\pm = J_1 \pm i J_2,\ee
then $[J_3, J_\pm] = \pm J_\pm$, and $[J_+, J_-] = 2J_3$.  Given any
complex number $z$, it is possible to find functions $\alpha(z, \bar{z})$
and $\beta(z, \bar{z})$, with $\beta(z, \bar{z})$ real, such that
\bge
e^{z J_-} e^{\azz J_+} e^{\bzz J_3} \in G.\label{eq:sucpx}\ee
The functions $\azz$ and $\bzz$ can be calculated explicitly by
working in the fundamental representation of $SU(2)$; one finds that
\begin{eqnarray}
\azz & = & \frac{-\bar{z}}{1 + |z|^2}, \label{eq:ablow} \\
\bzz & = & - \ln(1 + |z|^2). \nonumber \end{eqnarray}
Alternatively, these functions can be calculated in a
perturbative expansion about the identity, by applying the
Baker-Campbell-Hausdorff (BCH) formula \cite{Kir}
\bge
e^X e^Y = e^{X + Y + \frac{1}{2}[X,Y] + \ldots},\ee
which expresses the product of two exponentiated elements of a Lie
algebra in terms of a single exponentiated element of the algebra
as a formal power series.  (The ellipses in this formula denote
third- and higher-order commutators between $X$ and $Y$.)  The BCH
approach will be used in the next section to describe the complex
coordinate system we
will use on $\di$, so we will concentrate on this approach
here also, rather than using the more convenient exact expressions
which can be derived for $SU(2)$.

In some neighborhood of the identity, any element $g \in G$ can be
expressed uniquely in the form
\bge
g = e^{z J_-} e^{\azz J_+} e^{\bzz J_3 + i \psi J_3} \label{eq:gform} \ee
with $\psi$ real, so locally at least, $z$ is a good complex
coordinate on $G/H$.  To
see that the complex structure defined by $z$ is invariant under the
left action of $G$ on $G/H$, we multiply the above expression for $g$
on the left by an element $g' \in G$, also in the form (\ref{eq:gform}),
and get
\begin{eqnarray}
g' g & = & (g' e^{z J_-}) e^{\azz J_+} e^{\bzz J_3 + i \psi J_3} \nonumber \\
& = & (e^{z' J_-} e^{\alpha'' J_+} e^{\beta'' J_3}) e^{\azz J_+} e^{\bzz
J_3 + i \psi J_3}, \end{eqnarray}
where the BCH formula has again been applied several times, first to
rewrite $g' \exp(z J_-)$ as a single exponential, then again to
separate out the coefficients $\alpha''$ and $\beta''$.  By another round
of BCH-type manipulations, since $\{J_3, J_+\}$ generate a closed
subalgebra, we have
\bge  g' g = e^{z' J_-} e^{\alpha' J_+} e^{\beta' J_3 + i \psi' J_3}, \ee
for some coefficients $\alpha', \beta'$, and $\psi'$, with $\beta'$ and $\psi'$
real.  Since $g' g \in G$, we must have $\alpha' = \azzp$ and $\beta' = \bzzp$.
The value of $z'$ is purely a function of $g'$ and the holomorphic
coordinate $z$.  Thus, for fixed
$g'$, $z'$ is a holomorphic function of $z$, so the complex structure
defined by $z$ is invariant under left multiplication
by $g'$.  In fact, $z$ is just the usual complex coordinate on $S^2$
given by projection from the south pole onto $\CC$, which is
naturally invariant under the rotations generated by $SU(2)$.

{}From (\ref{eq:sucpx}) and (\ref{eq:ablow}), we can relate the
differentials $dz$, $d
\bar{z}$ to our original coordinates $b_i$.  At $\bbo$, we have
\begin{eqnarray}
dz & = & \frac{1}{2b} (db_1 + i db_2),  \\
d\bar{z} & = & \frac{1}{2b} (db_1 - i db_2). \nonumber \end{eqnarray}
It is now possible to express $\omega$ at $\bbo$ in terms of the
$z,\bar{z}$ coordinates; one finds that
\begin{eqnarray}
\omega_{\bar{z} z} & = & - \omega_{z \bar{z}} \; = \; 2bi,  \\
\omega_{z z} & = & - \omega_{\bar{z} \bar{z}} \; = \; 0. \nonumber
\end{eqnarray}
Thus, $\omega$ is indeed a (1,1)-form, and along with the
$G$-invariant complex structure given by $z$, defines a \kl structure
on $\wb$.  We can therefore restrict attention to the space $\hb$ of
holomorphic sections of $\lb$.  Since $\frac{- \omega}{2\pi}$ is the
first Chern class of $\lb$, for $b \geq 0$ it is a simple result of
the Riemann-Roch theorem
that $\lb$ admits exactly $2b+1$ linearly independent holomorphic
sections (see for example Griffiths and Harris \cite{GH}).
By choosing the proper local trivialization of
$\lb$, the $2b+1$ holomorphic sections are represented in the vicinity
of the origin $z=\bar{z}=0$ by the holomorphic monomials, $1, z, z^2,
\ldots, z^{2b}$.  $\lb$ also has a natural Hermitian metric, which we will
discuss further at the end of this section.

We now have a Hilbert space $\hb$, and there exist a set of operators
$\hat{J}_3,
\hat{J}_\pm$, given by
\bge
\hat{J}_a = \xi_a - A_a +i \Phi_a,\ee
for $a =3,\pm$, which act on $\hb$ to give a representation of
$\cg_{\CC}$.
We wish to compute these operators explicitly in terms of the complex
coordinate $z$.
First, we compute the vector fields $\tu_a$
corresponding to the actions of $\ads_{J_a}$.  (Technically, these are
vector fields in the complexification of the tangent space to $\wb$.)
Since we are only
concerned with the action of the differential operators associated
with these vector fields on holomorphic sections of $\lb$, which will
be written as holomorphic functions of $z$, it is only
necessary to compute the component of these vector fields in the
$\pdv/ \pdv z$ direction.  To compute the vector field at $z$ associated
with the coadjoint action of a generator $J_a \in \cg_{\CC}$, we must
express the product $\exp(\eps J_a) \exp (z J_-)$ in the form
\bge
e^{\eps J_a} e^{z J_-} = e^{(z + \eps \tu_a)J_-} f(J_3, J_+) +
\co(\eps^2) \ee
to first order in $\eps$, where $f(J_3, J_+)$ is some function of the
generators $J_3$ and $J_+$.  Since $\{J_3, J_+\}$ generate a closed
subalgebra of $\cg_{\CC}$, $\tu_a \pdv / \pdv z$ will be the tangent
vector to $\wb$ at $z$ associated with the action of $J_a$.  The
corresponding differential operator $\xi_a$ will then be defined by
$\xi_a = - \tu_a \pdv / \pdv z$.
To explicitly compute these operators, we will use the
infinitesimal forms of the BCH theorem (for a derivation of these
forms, see for example Kirillov \cite{Kir}),
\begin{eqnarray}
e^{\eps X} e^{Y} & = & \exp(Y + \eps \sum_{k \geq 0}
\frac{B_k}{k!} (\ad_Y)^k X) + \co(\eps^2), \label{eq:bch1} \\
e^{Y + \eps Z + \eps X} & = & \exp\left(Y + \eps Z - \eps \sum_{k \geq 1}
\frac{B_k}{k!} (-\ad_Y)^k X\right) \;\;e^{\eps X} + \co(\eps^2),
\label{eq:bch2} \end{eqnarray}
and
\bge
e^{\eps X} e^Y = e^{Y + \eps [X,Y]} e^{\eps X} + \co(\eps^2), \label{eq:bch3}
\ee
where $B_k$ is the $k$th Bernoulli number; $B_0 = 1, B_1 = -1/2, B_2 =
1/6, \ldots$.
Applying these formulae, we have
\begin{eqnarray}
e^{\eps J_3} e^{z J_-} & = & e^{(z - \eps z) J_-} e^{\eps J_3} +
\co(\eps^2),  \\
e^{\eps J_-} e^{z J_-} & = & e^{(z + \eps) J_-}, \nonumber
\end{eqnarray}
and
\begin{eqnarray}
e^{\eps J_+} e^{z J_-} & = & e^{z J_- + 2 \eps z J_3} e^{\eps J_+} +
\co(\eps^2)  \\
& = & e^{(z - \eps z^2) J_-} e^{2 \eps z J_3} e^{\eps J_+}  +
\co(\eps^2). \nonumber \end{eqnarray}
{}From these expressions, we can write the differential operators
$\xi_a$,
\begin{eqnarray}
\xi_3 & = & z \frac{\pdv}{\pdv z}, \nonumber \\
\xi_- & = & - \frac{\pdv}{\pdv z},  \\
\xi_+ & = & z^2 \frac{\pdv}{\pdv z}. \nonumber \end{eqnarray}
One can verify that these operators satisfy the proper commutation
relations.  We will now use the result of Proposition \ref{p:p2} to
construct the operators $\hat{J}_a$ explicitly, by choosing a
convenient form for the functions $f_a = - A_a +i \Phi_a$.  The operators
$\xi_a$ act on $\hb$, which in local coordinates is a subspace of the
ring $\CC[z]$ of polynomials in $z$.  $\CC[z]$ is a graded ring, with
a grading defined by $\deg (z^n) = n$.  The eigenvectors of $\xi_3$
are exactly the functions $z^n$ of fixed degree, with eigenvalues
equal to the degrees, so that $\xi_3 z^n = n z^n$.  A natural Ansatz
on the form of $f_3$ would be to insist that the operator $\hj_3 =
\xi_3 + f_3$ have the same eigenvectors as $\xi_3$.  This Ansatz
implies that $f_3$ is a constant function, and is equivalent to
performing a gauge fixing on $A_a$, given by $A_3 = - f_3 + i\Phi_3$.  A
priori, it is not obvious that this choice of gauge is possible, i.e.,
that this Ansatz is compatible with the conditions ({\sl i}) and ({\sl
ii$'$}) on the functions $f_a$ from Propositions \ref{p:p1} and
\ref{p:p2}.  We will proceed, however, to explicitly construct
functions consistent with both the Ansatz and these conditions.  In
fact, it turns out that the connection associated with this choice of
gauge is exactly the metric connection on $\lb$ associated with the
natural Hermitian structure.

The
constant $f_3$ is uniquely fixed to be $-b$ by condition ({\sl ii$'$}),
since $J_3$ stabilizes $\bbo$, and $\bbo(J_3) = -ib$.  From ({\sl i})
we have the relations
\bge
\xi_3 f_\pm - \xi_\pm f_3 = \pm f_\pm = \xi_3 f_\pm,\ee
so it follows that $f_- = 0$, and that $f_+$ is linear in $z$.  To fix
the coefficient of $f_+$, we use ({\sl i}) again, to show that
\bge
\xi_+ f_- - \xi_- f_+ = - \xi_- f_+ = \frac{\pdv f_+}{\pdv z} = 2 f_3
= -2b.\ee
Thus, we have constructed a set of functions
\bge
f_3 = -b, f_- = 0, f_+ = -2bz,\ee
which satisfy conditions ({\sl i}) and ({\sl ii$'$}) of Propositions
\ref{p:p1} and
\ref{p:p2}.  For every $b \in \ZZ/2$, then, we have explicitly
constructed a representation of $SU(2)$ on the space of polynomials in
$z$ of degree less than or equal to $2b$, given by the
operators
\begin{eqnarray}
\hj_3 & = & z \frac{\pdv}{\pdv z} - b, \nonumber \\
\hj_- & = & - \frac{\pdv}{\pdv z}, \label{eq:su2rep} \\
\hj_+ & = & z^2 \frac{\pdv}{\pdv z} - 2b z. \nonumber
\end{eqnarray}

To conclude this section, we will discuss briefly the unitary
structure of the coadjoint orbit representations of $SU(2)$ on $S^2$.
The rotationally invariant measure on $S^2$ in the coordinates we are using is
\bge
d \mu = \frac{2i}{(1 + |z|^2)^2} dz d \bar{z}.\ee
Along with this measure, there is a natural Hermitian metric on
$\lb$, given by
\bge
e^{h(z, \bar{z})} = \frac{1}{(1 + |z|^2)^{2b}}.\ee
Combining these factors, there is an inner product $\langle, \rangle$
on $\hb$ given by
\bge
\langle \phi, \psi \rangle = \int_{S^2} d \mu \: e^h \phi^* \psi =
\int_{S^2} \frac{2i dz d\bar{z}}{(1 + |z|^2)^{2b+2}} \phi^*(z) \psi(z),\ee
where $\phi$ and $\psi$ are arbitrary holomorphic sections of $\lb$.
Performing this integral explicitly, one finds that
\bge
\langle z^l, z^m \rangle = \delta_{l, m} {4 \pi}\left[(2b + 1)\left(
\begin{array}{c} 2b \\ l \end{array}
\right)\right]^{-1}.\label{eq:su2prod} \ee
This inner product on $\hb$ is of course proportional to the usual
inner product on unitary irreducible representation spaces of
$SU(2)$.  In fact, the inner product (\ref{eq:su2prod}) could have
been calculated up to a constant normalization factor by fixing
$\langle 1, 1 \rangle = 1$ and assuming that the representation of
$SU(2)$ on $\hb$ described by (\ref{eq:su2rep}) is unitary with
$J_3^{\dag} = J_3$ and  $J_-^{\dag} = J_+$.  This approach to calculating
an inner product on $\hb$ is equivalent to the one used in the
abstract algebraic construction of unitary $SU(2)$ representations.
It does not allow us to directly calculate the Hermitian metric on
$\lb$, but nonetheless provides $\hb$ with a Hilbert space structure.
In the case of $\di$, there is not a well-defined invariant measure on
the orbit space, and it will be necessary to use this indirect method
to put unitary structures on those representations which admit them.

We will now show that the connection $A$ on $\lb$ defined previously
by our gauge-fixing procedure is
precisely the metric connection on $\lb$ associated with the Hermitian
metric $\exp(h)$.  Recall that in general a Hermitian line bundle with
Hermitian
metric $\exp(h)$ has a metric connection given by $A_{\bar{z}} =
0$, $A_z = \pdv h / \pdv z$.  This is the unique connection compatible
with both the Hermitian metric and the complex structure \cite{GH}.  Thus, the
Hermitian connection on $\lb$ is given by
\bge
A_z = \frac{-2b \bar{z}}{1 + |z|^2}.\ee
In terms of $A_z$, the connection terms $A_a$, $a = 3, \pm$, are given
by
\begin{eqnarray}
A_3 & = & - z A_z \nonumber \\
A_+ & = & - z^2 A_z \\
A_- & = & A_z. \nonumber \end{eqnarray}
The connection $A_a = - f_a + i \Phi_a$ can be explicitly calculated
by evaluating
\begin{eqnarray}
\Phi_a(z) & = & - \langle \aads_{g(z, \psi)} \bbo, J_a \rangle \nonumber  \\
& = & - \langle \bbo, e^{- \azz J_+} e^{-z J_-} J_a e^{zJ_-} e^{\azz
J_+} \rangle. \end{eqnarray}
One finds that
\begin{eqnarray}
A_3 & = & - 2b z \azz, \nonumber \\
A_+ & = & - 2b z^2 \azz, \\
A_- & = & 2 b \azz. \nonumber \end{eqnarray}
This is exactly the Hermitian connection on $\lb$.

\section{$\di$ Virasoro representations}
\baselineskip 18.5pt

We will now turn our attention to the coadjoint orbits of the Virasoro
group, $\vir$.  The Virasoro group is the universal central extension
of the group of orientation-preserving diffeomorphisms of the circle,
$\dif$.  (For a clear discussion of central extensions and
infinite-dimensional groups, see \cite{PS}.)  Elements of $\vir$ are
given by pairs $(\phi, \alpha)$, with $\phi \in \dif$, and $\alpha \in
U(1)$.  The Lie algebra of $\vir$, which we denote $\vvec$, is
likewise the universal central extension of the algebra of smooth
vector fields on $S^1$.  Elements of $\vvec$ are of the form $(f, -i
a)$, with $f(\theta) \pdv/ \pdv \theta$ a vector field on $S^1$ and $a
\in \RR$.  (Except for a few signs, we mostly use the notation of
Witten \cite{Witt1} in this section.)  The commutation relation between
elements of $\vvec$ is given by
\bge
[(f, -ia_1), (g, -ia_2)] = \left( f g' - gf', \frac{i}{48 \pi} \int_0^{2
\pi} (f(\theta) g'''(\theta) - g(\theta) f'''(\theta)) d \theta \right).\ee
Defining the (complex) vector fields $l_n = i e^{i n \theta} \pdv /
\pdv \theta$ in $\vc$, we can define the usual Virasoro generators
by
\begin{eqnarray}
L_n & = & (l_n, 0); \;\;{\rm for} \; n \neq 0, \nonumber \\
L_0 & = & (l_0, \frac{1}{24}), \label{eq:vgen} \\
C & = & (0, 1). \nonumber \end{eqnarray}
The commutation relations then take the standard form
\begin{eqnarray}
[L_m, L_n] & = & (m - n) L_{m + n} + \frac{C}{12}(m^3 - m) \delta_{m
, - n},  \\
\lbrack C, L_n \rbrack & = & 0. \nonumber \end{eqnarray}
The Virasoro algebra is defined to be the complex Lie algebra spanned
by the generators (\ref{eq:vgen}).  The (smooth) dual space to $\vvec$
consists of pairs $(b, it)$, with $b(\theta) d \theta^2$ a quadratic
differential on $S^1$, and $t \in
\RR$.  The dual pairing between $(b, it)$ and an element $(f, -ia) \in
\vvec$ is given by
\bge
\langle (b, it), (f, -ia) \rangle = \int_0^{2 \pi} b(\theta) f(\theta)
d \theta + at.\ee
For this pairing to be invariant under the action of the algebra
$\vvec$, $(b, it)$ must transform under the coadjoint action by
\bge
\ads_{(f, -ia)} (b, it) = (2 b f' + b' f - \frac{t}{24 \pi} f''',
0).\ee
By computing the stabilizer of a general dual element $(b, it)$, it
is possible to completely classify the coadjoint orbits of $\vir$.  A
clear review of this analysis in the general case is given in
\cite{Witt1}.  We will only be concerned here with the simplest case,
in which the orbit contains an element $(b_0, ic)$ with $b_0(\theta) =
b_0$ a constant function.  We will refer to this orbit as $\wbc$.  In
this case, the stabilizer in $\vvec$ of the point $(b_0, ic)$ is given
by all elements $(f, -ia)$ with $f(\theta)$ satisfying
\bge
\frac{c}{24 \pi} f''' = 2 b_0 f'.\ee
When $- 48 \pi \frac{b_0}{c}$ is not the square of an integer $n$, the
only solution to this equation with period $2\pi$ is $f(\theta) = 1$.
In this case,
the stabilizer of $\bc$ is the subgroup generated by $L_0$ and $C$, so
the space $\wbc$ is equivalent to the space $\di$.  For the
exceptional values of $b_0, c$, the generators $\L_{\pm n}$ are also
stabilizers of $\bc$.  Thus, the coadjoint orbits $W_{-cn^2/ 48\pi,
c}$ are given by the spaces $\dis$, where $SL^{(n)}(2, \RR)$ is
generated by the elements $l_0, l_{\pm n}$ in $\vc$.  We will not
concern ourselves here with the orbits $W_{-cn^2/ 48\pi,c}$, but we
will find that even when $b_0 = - \frac{c n^2}{48 \pi}$, a
representation of the Virasoro group can be constructed on the space
$\di$.

{}From now on, we will consider a fixed orbit $\wbc$, of the $\di$ type.
The procedure we will follow in constructing the Virasoro
representation corresponding to $\wbc$ is exactly that which we
followed for $SU(2)$ in the previous section.  First, we must
evaluate $\omega$ at a fixed point in $\wbc$, which we choose to be
$\bc$.  Then, after checking the integrality condition, which is
trivial in this case since the space $\di$ is retractable, we will
construct suitable coordinates on $\di$ with respect to which $\omega$
is a (1,1)-form.  In this coordinate system, we will explicitly
calculate the form of the operators $\hln$, using the result of
Proposition \ref{p:p2}.  Let us proceed to evaluate $\omega$ at the
point $\bc$.  A basis for the tangent space to $\wbc$ at $\bc$ is
given by the real vector fields corresponding to the generators of
$\vir$, $\{L_n - L_{-n}, i (L_n + L_{-n}) : n \geq 1\}$.  We could calculate
the components of $\omega$ with respect to this basis; it will be most
convenient, however, to simplify our notation by computing directly
the complex linear combinations of those components given by
\begin{eqnarray}
\omega_{m,n} & = & \langle \bc, [L_m, L_n] \rangle \nonumber \\
& = & i \delta_{m, -n} (4 \pi m b_0 + \frac{c}{12} m^3). \label{eq:omvir}
\end{eqnarray}
This  2-parameter family of symplectic
structures is in fact the most general form for an invariant 2-form on $\di$.
Note that when $\bo = - \frac{c n^2}{48 \pi}$, the 2-form
$\omega$ is degenerate, and thus is not a symplectic form.

We now make the observation that $\di$ is a contractible space.  To
see this, note that $\di$ can be identified with the group $\dio$ of
orientation-preserving diffeomorphisms of $S^1$ which fix the point 1.
Viewing an element of $\dio$ as a monotonically increasing function $f: \RR
\rightarrow \RR$ with the properties $f(0) = 0$ and $f(x + 2 \pi)
= 2 \pi + f(x)$, we can explicitly give a
retraction of $\dio$ to a point by defining the one-parameter family
of functions $f_t(x) = (1-t) f(x) + tx$, for each $f \in \dio$, $t \in
[0,1]$.  Since $\di$ is a contractible space, all 2-cycles are
homologous to the null 2-cycle, so that $\int_a \omega = 0$ for any
2-cycle $a$.  Thus, for any $b_0, c$, we can construct a line bundle
$\lbc$ over $\wbc$ with curvature $i \omega$.  (Note that the second cohomology
of $\di$ is nontrivial if one restricts to forms invariant under
$\dif$, however this should not affect the construction of $\lb$; it
does however imply that $\lb$ will not have a global $\dif$-invariant
connection.)

The next step in our
construction is to give a complex structure to $\wbc$ with
respect to which $\omega$ is a (1,1)-form.  This complex structure can
be constructed in a fashion similar to the one used for $SU(2)$ in the
previous section.  Given a countable set of variables $z = \{z_1, z_2,
\ldots\}$, there exist unique functions $\mzz, \rzz, \gzz$, expressed as
formal power series in the $z_i's$, such that $\rzz$ and $\gzz$ are
real and
\bge
\exp (\sum_{n>0} z_n L_n) \exp(\sum_{n>0} \mzz L_{-n}) \exp(\rzz L_0)
\exp(\gzz C) \in \dio. \label{eq:zcoord} \ee
As in the case of $SU(2)$, the functions $\mu_n, \rho$, and $\gamma$
can be explicitly calculated order-by-order in the $z$'s by applying
the BCH formula.  Up to second order terms in the $z$'s, these
functions are given by
\begin{eqnarray}
\mzz & = & - \bar{z}_n + \sum_{m>0} (n+2m)z_m \bar{z}_{n+m} +
\co(z^3),\nonumber
\\ \rzz & = & \sum_{k>0} k |z_k|^2 + \co(z^3), \\
\gzz & = & \sum_{k>0} \frac{k^3 - k}{24} |z_k|^2 + \co(z^3).
\nonumber \end{eqnarray}
The complex variables defined by (\ref{eq:zcoord}) were successfully
used in a previous work of Zumino \cite{Zum} to calculate the
curvature of $\di$.  (This curvature calculation was first done by
Bowick and Rajeev using other methods \cite{BR}.)  Nevertheless, the
unusual geometry of the Virasoro group (specifically the fact that the
exponential map is not locally 1-1 or onto) calls the validity of this
coordinate system into question; thus, we will briefly outline an
argument justifying this choice of coordinate system.  We will not
attempt to be mathematically rigorous here; work is currently underway to
provide a mathematically complete justification for this point of
view.  The basic point is that the exponential map on $\dif$ fails to
be well-behaved due to diffeomorphisms which are either non-analytic,
or contain no fixed points \cite{Milnor,PS}.  By restricting to
$\dio$, we eliminate the latter problem.  If we also restrict attention
to the subgroup of $\dio$ consisting of diffeomorphisms which are
real-analytic, the problems with the exponential map relating to
non-analytic diffeomorphisms are also
removed.  There are several reasons that it is reasonable to restrict
attention to real-analytic diffeomorphisms.  The first is that in most
cases of physical interest, non-analytic maps are not of concern; in
conformal field theory, for instance, conformal transformations on the
world-sheet which leave a time-slice fixed are real-analytic
diffeomorphisms on that time-slice.  The second reason for allowing the
restriction to real-analytic diffeomorphisms is that Goodman and
Wallach have shown \cite{GW}, following a conjecture of Kac \cite{Kac},
that every unitary representation of the algebra of real-analytic
vector fields on $S^1$ can be integrated to a continuous unitary
representation of $\vir$.  (Actually, their proof assumes only that a
representation can be found for the algebra of vector fields with
finite Fourier series.)

We will assume that for the group $D_0$ of real-analytic
diffeomorphisms in $\dio$, the exponential map from the associated
algebra is 1-1 and onto.  We do not yet have a rigorous proof of this
assertion, however it is not hard to show that the exponential map is
1-1 and onto in the closely related case where we consider the group
of all formal power series in one variable with first nonvanishing
term positive and linear, under the group law given by composition of
functions.
With this assumption, we have a set of coordinates $\beta =
\{\beta_1, \beta_2, \ldots\}$ on $D_0$, given by
\bge
g(\beta) = \exp\left[\sum_{n > 0} (\beta_n (L_n - L_0) -
\bar{\beta}_n(L_{-n} - L_0))\right].\ee
By using the BCH theorem, these coordinates and the coordinates $z$
defined by (\ref{eq:zcoord}) can be expressed in terms of one another
as formal power series; this gives in a formal sense a 1-1 correspondence
between the coordinates $z$ and $\beta$, so that we can consider $z$
to be a global coordinate system on $D_0 \sim \di$.

{}From (\ref{eq:omvir}), it is clear that in the $z$ coordinates the
curvature form $\omega$ is given at the origin $z_n = 0$ by
\begin{eqnarray}
\omega_{\bar{m},n} & = & - \omega_{n, \bar{m}} = i \delta_{m,n} (4\pi
m \bo + \frac{c}{12} m^3)  \\
\omega_{m,n} & = & \omega_{\bar{m}, \bar{n}} = 0. \nonumber
\end{eqnarray}
Thus, $\omega$ is a (1,1)-form, and we have a \kl structure on $\wbc$.
In the coordinate system given by $z$, the space of
holomorphic sections of $\lbc$ can be taken to be the ring $R = \CC
[z_1, z_2, \ldots]$ of polynomials in the variables $z_i$.  We now
wish to explicitly compute the action of the operators
\bge
\hln = \xn - A_n + i \pn,\ee
which give a representation of the Virasoro algebra
on $R$.  Note that there is also an operator $\hat{C} = c$,
which is constant since $C$ is central.  From now on we will simply
replace the operator $\hat{C}$ with its value $c$ in all formulae.

We begin the computation of the $\hln$'s, as in the case of $SU(2)$,
by computing the vector fields $\xn$.  Expressions for these vector
fields can be calculated by using BCH to express the product $\exp
(\eps L_n) \exp( \sum z_m L_m)$ in the form
\bge
\exp (\eps L_n) \exp(\sum_{m > 0} z_m L_m) = \exp(\sum_{m > 0} (z_m +
\eps \tu_n^m) L_m) \;f(\{L_k: k \leq 0\}, C) + \co(\eps^2),
\ee
and then setting
\bge
\xn = \sum_{m > 0} - \tu_n^m \frac{\pdv}{\pdv z_m}.
\ee
\begin{prop}
The vector fields $\xn$ are given by
\bge
\xn = \sum_{\sss k \geq 0, N_n^+ (k)}
\alpha_{k,\lambda} C_n(n_1, \ldots, n_k) z_{n_1} \ldots z_{n_k}
\frac{\pdv}{\pdv
z_{n+n_1+\ldots + n_k}}, \label{eq:xiform} \ee
where $\lambda$ is the minimum integer
such that $n + n_1 + n_2 + \ldots + n_{\lambda} > 0$ {\rm (}$\lambda = 0$
when $n > 0${\rm )},
\bge
N_n^{\pm} (k) = \{(n_1, n_2, \ldots, n_k) : n_1,\ldots,n_k > 0 ,n +
n_1 + n_2 + \ldots + n_k
\begin{array}{cc}> \\ \leq \end{array} 0\},\ee
\bge C_n(n_1, \ldots, n_k) = (n_1 - n)(n_2 - n_1 - n)
\ldots (n_k - n_1 -\ldots - n_{k-1} - n), \ee
and
\bge \alpha_{k, \lambda} = (-1)^{k+1} \sum_{l=0}^{k-\lambda}
\frac{B_l}{l! \: (k-l)!}.\ee
{\rm (}$B_l$ is the $l$th Bernoulli number, as in (\ref{eq:bch1}) and
(\ref{eq:bern}).{\rm )}
\label{p:xcalc}
\end{prop}
\noindent {\it Proof.} \,\,
We begin by noting the identities
\bge
(\ad_{(\sum z_m L_m)})^k L_n = \sum_{\sss n_1, \ldots, n_k >
0} C_n(n_1, \ldots, n_k) z_{n_1} \ldots z_{n_k} L_{n + n_1 + \ldots +
n_k},\label{eq:adk} \ee
and
\bge C_n(n_1, \ldots, n_s) C_{n+ n_1 + \ldots + n_s}(n_{s+1}, \ldots, n_k)
= C_n(n_1, \ldots, n_k). \label{eq:ccomb} \ee
Equation \ref{eq:adk}
also contains a constant term proportional to $C$ on the right hand
side.  Since $C$ is central, this term can immediately be absorbed in
$f$, and will be dropped from all calculations in this proof.
Applying Equation \ref{eq:bch1} to $\exp(\eps L_n) \exp(\sum
z_m L_m)$, we have
\begin{eqnarray}
\lefteqn{\exp(\eps L_n) \exp(\sum_{m > 0} z_m L_m) \sim} \nonumber \\
& & \exp \left(\sum_{m > 0} z_m L_m
+ \eps \sum_{\sss k \geq 0, n_1, \ldots, n_k > 0} \frac{B_k}{k!}
C_n(n_1, \ldots, n_k) z_{n_1} \ldots z_{n_k} L_{n + n_1 + \ldots +
n_k}\right), \end{eqnarray}
where by $a \sim b$ it is meant that $a = bf + \co(\eps^2)$, with $f$
some function of the $L_n$'s with $n \leq 0$.
Dividing the terms in the exponential into generators $L_m$ with $m >
0$ and $m \leq 0$, this can be rewritten as
\begin{eqnarray}
\lefteqn{\exp(\eps L_n) \exp(\sum_{m > 0} z_m L_m) \sim} \nonumber
\\ & &  \exp \left(\sum_{m > 0} z_m L_m
- \eps \sum_{\sss k \geq 0, N_n^+(k)} \alpha_{k,\lambda}^{(0)}
C_n(n_1, \ldots, n_k) z_{n_1} \ldots z_{n_k} L_{n + n_1 + \ldots +
n_k} \right.\nonumber \\ & & \left.\;\;\;\;\;\;\; + \eps \!\!\!
\sum_{\sss l_1 \geq 0, N_n^-(l_1)}
\! \frac{B_{l_1}}{l_1 !} C_n(n_1, \ldots, n_{l_1}) z_{n_1} \ldots
z_{n_{l_1}} L_{n + n_1 + \ldots + n_{l_1}}\right), \label{eq:exp0}
\end{eqnarray}
 where
$\alpha_{k, \lambda}^{(t)}$ is defined by
\bge \alpha_{k, \lambda}^{(t)} = - \frac{B_k}{k!} - \!\!\! \sum_{\sss
t \geq s > 0, 0 \leq l_1
< l_2 < \ldots < l_s < \lambda} \!\!\! (-1)^{s + k - l_1}
\frac{B_{l_1}}{l_1 !} \frac{B_{l_2 - l_1}}{(l_2 - l_1) !} \ldots
\frac{B_{l_s - l_{s-1}}}{(l_s - l_{s-1}) !} \frac{B_{k - l_s}}{(k -
l_s) !}. \ee
Applying (\ref{eq:bch2}) and (\ref{eq:ccomb}) to Equation
\ref{eq:exp0} $t$ times, we get
\begin{eqnarray}
\lefteqn{\exp(\eps L_n) \exp(\sum_{m > 0} z_m L_m)  \sim} \nonumber \\
& & \exp \left(\sum_{m > 0} z_m L_m
- \eps \sum_{\sss k \geq 0, N_n^+(k)} \alpha_{k,\lambda}^{(t)}
C_n(n_1, \ldots, n_k) z_{n_1} \ldots z_{n_k} L_{n + n_1 + \ldots +
n_k} \right. \\ & & \left. \;\;\;\;\;\;\; + \eps \!\!\!\!\!\!
\sum_{\sss 0 \leq l_1 < \ldots <
l_t < l, N_n^-(l)} \!\!\!\!\! (-1)^{t+l-l_1}
\frac{B_{l_1}}{l_1 !}
\frac{B_{l_2 - l_1}}{(l_2 - l_1) !} \ldots
\frac{B_{l - l_{t}}}{(l - l_{t}) !}
C_n(n_1, \ldots, n_{l}) z_{n_1} \ldots z_{n_{l}} L_{n + n_1 + \ldots +
n_{l}}\right). \nonumber \end{eqnarray}
Since $\alpha_{k, \lambda}^{(t)} =
\alpha_{k,
\lambda}^{(\infty)}$ for $t \geq \lambda$, to all orders in $z$ we
have
\begin{eqnarray}
\lefteqn{\exp(\eps L_n) \exp(\sum_{m > 0} z_m L_m)  \sim} \nonumber \\
& & \exp \left(\sum_{m > 0} z_m L_m
- \eps \sum_{\sss k \geq 0, N_n^+(k)} \alpha_{k,\lambda}^{(\infty)}
C_n(n_1, \ldots, n_k) z_{n_1} \ldots z_{n_k} L_{n + n_1 + \ldots +
n_k} \right), \end{eqnarray}
We will now show that $\alpha_{k,\lambda}^{(\infty)} =
\alpha_{k,\lambda}$.  Using the fact that $B_{2k+1} = 0$ for $k >
0$, it is not hard to determine that
\begin{eqnarray}
\alpha_{k, 0}^{(\infty)} & = & - \frac{B_k}{k!}, \nonumber \\
\alpha_{k, 1}^{(\infty)} & = & \delta_{k, 1}, \label{eq:alkl} \\
\alpha_{k, \lambda}^{(\infty)} & = & \!\! \sum_{\sss s \geq 0, 1 < l_1
< \ldots < l_s <
\lambda} \!\! (-1)^{s+k}
\frac{B_{l_1 - 1}}{(l_1 - 1) !} \frac{B_{l_2 - l_1}}{(l_2 - l_1) !}
\ldots \frac{B_{l_s - l_{s-1}}}{(l_s - l_{s-1}) !}
\frac{B_{k - l_s}}{(k - l_s) !}, \;\;{\rm for}\; \lambda > 1.
\nonumber \end{eqnarray}
When $\lambda > 1$, we can write a generating function for $\alpha_{k,
\lambda}^{(\infty)}$ by
\bge
\sum_{k \geq \lambda > 1}  \alpha_{k,\lambda}^{(\infty)} y^{k-\lambda}
x^k = \sum_{l > m \geq 0} y^m \frac{B_l}{l!} (-x)^{l+1} \sum_{s \geq 0}
(1-\phi(-x))^s, \label{eq:gen} \ee
where
\bge
\phi(x) = \frac{x}{e^x - 1} = \sum_{n \geq 0} \frac{B_n}{n!}
x^n.\label{eq:bern} \ee
{}From (\ref{eq:gen}), it follows that
\bge \alpha_{k, \lambda}^{(\infty)} = (-1)^k \sum_{l=k - \lambda + 1}^{k-1}
\frac{B_l}{l! \: (k-l)!} = (-1)^{k+1} \sum_{l=0}^{k-\lambda}
\frac{B_l}{l! \: (k-l)!},\ee
where we have used the fact that
\bge
\sum_{l=0}^{k-1} B_l \left( \begin{array}{cc} k \\ l \end{array}
\right) = 0,\;\;\;{\rm
for}\; k > 1. \label{eq:bid} \ee
{}From (\ref{eq:alkl}) and (\ref{eq:bid}), it is also easy to verify that
$\alpha_{k,0}^{(\infty)} = \alpha_{k,0}$ and $\alpha_{k,1}^{(\infty)}
= \alpha_{k,1}$.  Thus, for all $k$ and $\lambda$, we have shown that
$\alpha_{k,\lambda}^{(\infty)} = \alpha_{k,\lambda}$, and
proposition \ref{p:xcalc} is proven.$\Box$

When $n \geq 0$, we can use (\ref{eq:alkl}) to simplify the formulae
for $\xn$
to
\bge
\xn = \!\sum_{\sss k \geq 0, n_1, \ldots, n_k > 0}\!
 - \frac{B_k}{k!} C_n(n_1, \ldots, n_k) z_{n_1} \ldots z_{n_k} \frac{\pdv}{\pdv
z_{n+n_1+\ldots + n_k}}\;\;\; {\rm for} \; n > 0,\ee
and
\bge
\xi_0 = \sum_{k > 0} k z_k \frac{\pdv}{\pdv z_k}.\ee
(Note that the last expression could also have been obtained more
simply through (\ref{eq:bch3}).)
The first few terms in the expressions for $\xn, n \neq 0$ are given
by
\bge
\xi_n = - \frac{\pdv}{\pdv z_n} + \frac{1}{2} \sum_{m > 0} (m-n) z_m
\frac{\pdv}{\pdv z_{n+m}} + \ldots, \;\;\; {\rm for} \; n > 0,\ee
and
\bge
\xi_{-n} = \sum_{m > n} (m+n) z_m
\frac{\pdv}{\pdv z_{m-n}} + \ldots, \;\;\; {\rm for} \; n > 0.\ee
It is important to note that although the operators $\xn$ are
expressed as infinite series, to compute $\xn f$ for any $f \in R$ and
$n \in \ZZ$,
only a finite number of terms in $\xn$ will be needed, so that the
action of $\xn$ on any element of $R$ can be explicitly computed.

To calculate the operators $\hln$, it will be sufficient, by
Proposition \ref{p:p2}, to find a set of functions $f_n(z_1, z_2,
\ldots) \in R$, for $n \in \ZZ$, with the properties
\bge
\xi_m f_n - \xi_n f_m = (m - n) f_{m + n} + \frac{c}{12} \delta_{m,
-n} (m^3 - m) \label{eq:calpha} \ee
and
\bge
f_0 (0) = 2 \pi b_0 + \frac{c}{24}. \label{eq:cbeta} \ee
As in the $SU(2)$ example, we will find these functions by making an
Ansatz which will be justified once we find a set of $f$'s satisfying
(\ref{eq:calpha}) and (\ref{eq:cbeta}).  The carrier space $R$ is
again a graded ring, with $\deg(z_n) = n$, $\deg(1) = 0$.  The
eigenvectors of the $\xi_0$ operator are exactly those polynomials in
$R$ of fixed degree with respect to this grading; we refer to such
polynomials as {\it quasi-homogeneous}.  We can express $R$ as a
direct sum of finite dimensional vector spaces,
\bge
R = \bigoplus_{k \geq 0} R_k,
\ee
where
\bge  R_k = \{f \in R : \deg(f) = k\}. \ee
With this notation, we have
\bge
\xi_m R_n \subset R_{n-m}\ee
and
\bge  \xi_0 f = n f, \;{\rm for}\; f \in R_n.\ee
A natural Ansatz to make is that $\hat{L}_0$
satisfies
\bge  \hat{L}_0 R_n \subset R_n, \ee
which is equivalent to the assertion that $f_0$ is a constant
function.  In fact, we will show that this Ansatz leads to a unique
set of functions $f_n$ which satisfy (\ref{eq:calpha}) and
(\ref{eq:cbeta}).  An immediate result of the Ansatz is that
\begin{eqnarray}
f_n & = & 0, \;{\rm for}\; n > 0,  \\
f_{-n} & \in & R_n, \;{\rm for}\; n > 0,  \nonumber
\end{eqnarray}
so that $f_{-n}$ is a quasi-homogeneous function of degree $n$.  This
follows from setting $m = 0$ in (\ref{eq:calpha}).  We can easily
calculate the first few $f$'s by hand, using (\ref{eq:calpha}) and
(\ref{eq:cbeta}).  Defining $h = 2 \pi b_0 + \frac{c}{24}$, we have
{}from (\ref{eq:cbeta}),
\bge
f_0 = h. \label{eq:f0}
\ee
{}From (\ref{eq:calpha}), we have
\bge
\xi_1 f_{-1} = - \frac{\pdv}{\pdv z_1} f_{-1} = 2 f_0 = 2h,\ee
so
\bge
f_{-1} = -2h z_1. \label{eq:f1} \ee
Similarly, (\ref{eq:calpha}) gives two equations for $f_{-2}$,
\begin{eqnarray}
\xi_2 f_{-2} & = & - \frac{\pdv}{\pdv z_2} f_{-2} = 4 f_0 + \frac{c}{2} = 4h
+ \frac{c}{2},  \\
\xi_1 f_{-2} & = & - \frac{\pdv}{\pdv z_1} f_{-2} = 3 f_{-1} = -6 h z_1.
\nonumber \end{eqnarray}
Since $f_{-2}$ is a linear combination of $z_1^2$ and $z_2$, these two
equations determine both coefficients exactly, so that
\bge
f_{-2} = - (4h + \frac{c}{2}) z_2 + 3 h z_1^2. \label{eq:f2} \ee
One could continue computing the functions $f_{-n}$ by this means,
however the number of conditions on each function grows faster than
the number of linearly independent terms which can appear in the
same function.  Thus, it is desirable to find a means of expressing the
functions $f_{-n}$ for $n > 2$ in such a way that the consistency of
these conditions can be easily verified.  In fact, we can use
(\ref{eq:calpha}) to give a recursive definition of $f_{-n}$ in terms
of $f_{-1}$ and $f_{1-n}$.  We define
\bge
f_{-n} = \frac{1}{n-2} (\xi_{-1} f_{1-n} - \xi_{1-n} f_{-1}), \;{\rm
for}\; n > 2. \ee
We must now prove that this definition, along with Equations
\ref{eq:f0}, \ref{eq:f1}, and \ref{eq:f2}, gives a set of functions
$f_{-n}$ which are consistent with (\ref{eq:calpha}).  As an
intermediate result, we will need the following proposition.
\begin{prop}
The unit element 1 in $R$, which we denote by $\nul$, is the unique
function in $R$ (up to scalar multiplication) which is annihilated by
$\xi_n$ for all $n > 0$; i.e., $\nul$ is the unique highest weight
state in the module $R$ under the action of the $\xn$'s.
\label{p:p4}
\end{prop}
\noindent {\it Proof.} \,\,
Assume there is another function $\phi \in R$ which is annihilated by
$\xn$ for all $n > 0$.  Since $R$ is graded, $\phi$ can be written as
a sum of quasi-homogeneous functions,
\bge
\phi = \sum_{n \geq 0} \phi_n, \;\;\; \phi_n \in R_n.\ee
Take $d$ to be the minimum integer with $\phi_d \neq 0$.  Now, let $k$
be the largest integer such that some term in $\phi_d$ contains a
factor of $z_k$.  $\phi_d$ can now be written in the form
\bge
\phi_d = \sum_{m \geq 0} z_k^m g_{d-km}^{(m)} (z_1, \ldots, z_{k-1}),\ee
where $g_{d-km}^{(m)}$ is a quasi-homogeneous polynomial in $z_1,
\ldots, z_{k-1}$ of degree $d-mk$ for each $m \geq 0$.  Since for $n >
0$, all terms in $\xn$
except the leading term $- \pdv / \pdv z_n$ contain derivatives $\pdv
/ \pdv z_{j}$ with $j > n$, we can compute
\bge
\xi_k \phi_d = - \sum_{m > 0} m z_k^{m-1} g_{d - km}^{(m)} (z_1,
\ldots, z_{k-1}).\ee
For this expression to be zero, all the functions $g_{d-km}^{(m)}$
would have to  be zero for $m > 0$.  But then $\phi_d$ would not
contain any terms
with a factor of $z_k$, contradicting our assumption.  Thus, the only
states in $R$ annihilated by all $\xi_n$ with $n > 0$, are the constant
functions in $R_0$.$\Box$

{}From this proposition, it is clear that $h$ is in fact the standard
value of the highest weight for the Virasoro representation we are
constructing, since
$\hat{L}_0 \nul = f_0 \nul = h \nul$.  Note also that the proof of
Proposition \ref{p:p4} shows that there are not even any formal power
series in $\CC[[z_1, z_2, \ldots]]$ annihilated by $\xn$ for all $n >
0$, other than $1$.  ($\CC[[z_1, z_2, \ldots]]$ is the ring of formal
power series in the variables $z_1, z_2, \ldots$ with coefficients in
$\CC$, which is also a module under the action of the $\xi_n$'s.)
We can now show that the functions $f_{-n}$ defined previously do in
fact satisfy (\ref{eq:calpha}).
\begin{prop}
The functions $f_n \in R$ given by the recursive formula
\begin{eqnarray}
f_n & = & 0, \;\;\; n > 0, \nonumber \\
f_0 & = & h, \nonumber \\
f_{-1} & = & -2 h z_1, \label{eq:fdef} \\
f_{-2} & = & -(4h + \frac{c}{2}) z_2 + 3 h z_1^2, \nonumber \\
f_{-k} & = & \frac{1}{k-2} (\xi_{-1} f_{1-k} - \xi_{1-k} f_{-1}), \;\;\; k
> 2,\nonumber \end{eqnarray}
satisfy (\ref{eq:calpha}), where the operators $\xi_n$ are given by
Equation \ref{eq:xiform}.
\label{p:p5}
\end{prop}
\noindent {\it Proof.} \,\,
We divide the equations derived from (\ref{eq:calpha}) into two
categories, depending on whether the indices $m$ and $n$ are both
negative, or are
of opposite sign.  (When $m$ and $n$ are both positive, or $m$ or $n$
is zero, (\ref{eq:calpha}) is easily verified directly from the
definition of the $f$'s.)  When $m$ is positive, (\ref{eq:calpha})
states that for every $k > 0$,
\bge
\xi_m f_{-k} = (m + k) f_{m-k} + \frac{c}{12} \delta_{m,k} (m^3 - m)
\;\;\; {\rm for} \; {\rm all} \; m > 0. \label{eq:mpos} \ee
On the other hand, when the indices are both negative,
(\ref{eq:calpha}) states that for every $k > 0$,
\bge
\xi_{-l} f_{l-k} - \xi_{l-k} f_{-l} = (k - 2l) f_{-k}\;\;\; {\rm for} \;
{\rm all} \; l: 0 < l < k. \label{eq:mneg} \ee
We will prove by induction on $k$ that Equations \ref{eq:mpos} and
\ref{eq:mneg} follow from the above definitions of the functions
$f_n$.  From the discussion leading to Equations \ref{eq:f0},
\ref{eq:f1}, and \ref{eq:f2}, it is clear that for $k \leq 2$,
(\ref{eq:mpos}) is satisfied.  (Note that $\xi_m
f_{-k} = 0$ whenever $m > k \geq 0$, since all terms in $\xi_m$
contain derivatives with respect to some $z_j$ with $j \geq m$.)  It is
also easy to verify (\ref{eq:mneg}) for $k \leq 2$.  Thus, we have the
first induction steps; it remains to be shown that if (\ref{eq:mpos})
and (\ref{eq:mneg}) are satisfied for all $k' < k$, then they are also
satisfied for $k$.  We claim that to prove this, it will suffice to
show that for all $m >0$ and $0 < l < k$,
\bge
\xi_m \left[\frac{1}{k-2l} (\xi_{-l} f_{l-k} - \xi_{l-k} f_{-l})\right] = (m+k)
f_{m-k} + \frac{c}{12} \delta_{m,k} (m^3 - m). \label{eq:induct} \ee
Certainly, verifying this equation for $l = 1$ will show that
(\ref{eq:mpos}) holds for $k$.  In fact, however,
this equation also tells us that for any $l \neq 1$,
\bge
\xi_m \left[f_{-k} - \frac{1}{k-2l} (\xi_{-l} f_{l-k} - \xi_{l-k}
f_{-l})\right] = 0
\;\;\; {\rm for} \; {\rm all} \; m > 0.\ee
By Proposition \ref{p:p4}, this implies that
\bge
f_{-k} - \frac{1}{k-2l} (\xi_{-l} f_{l-k} - \xi_{l-k} f_{-l}) \in \CC
\;\;\; {\rm for} \; {\rm all} \; m > 0.\ee
Since the left hand side of this equation is quasi-homogeneous of
degree $k$, for $k > 2$, (\ref{eq:mneg}) follows.  Thus, to prove the
induction step, it will suffice to prove (\ref{eq:induct}).  The
derivation of this equation is straightforward algebra, assuming that
(\ref{eq:calpha}) and
(\ref{eq:cbeta}) hold for $k' < k$.
\begin{eqnarray}
\xi_m[\frac{\xi_{-l} f_{l-k} - \xi_{l-k} f_{-l}}{k-2l}] & = &
\frac{1}{k-2l} \left[ (m+l) \xi_{m-l} f_{l-k} + \frac{c}{12} \delta_{m,l}
(m^3 - m) f_{l-k} + \xi_{-l} \xi_m f_{l-k} \right. \nonumber \\ & &
\left. - (m + k - l) \xi_{m + l -
k} f_{-l} - \frac{c}{12} \delta_{m, k-l} (m^3 - m) f_{-l} - \xi_{l-k}
\xi_m f_{-l} \right] \nonumber \\
& = & \frac{1}{k-2l} \left[ (m+l) (\xi_{m-l} f_{l-k} - \xi_{l-k}
f_{m-l}) \right. \nonumber \\ & &
\left. - (m + k - l) (\xi_{m + l - k} f_{-l} - \xi_{-l} f_{m + l - k})
\right] \nonumber \\
& = & (m+k) f_{m-k} + \frac{c}{12} \delta_{m,k} (m^3 - m).
\end{eqnarray}
We have thus shown by induction that (\ref{eq:mpos}) and
(\ref{eq:mneg}) hold for all $k > 0$, so the proposition is proven.$\Box$

We have defined functions $f_n$ which satisfy (\ref{eq:calpha})
and (\ref{eq:cbeta}), and thus by Proposition \ref{p:p2} we have
constructed a representation of the Virasoro algebra on $R$, where the
generators
are given by
\begin{eqnarray}
\hln & = & \xn + f_n  \\
\hat{C} & = & c. \nonumber \end{eqnarray}
Note that the explicit formulae for $\xn$ and $f_n$ can be used to
construct a representation of $\vir$, at least formally, even when $h
= \frac{-c(m^2-1)}{24}$ for some integer $m$.  The action of each
operator $\hat{L}_n$ on any polynomial in $R$ involves a finite number
of terms,
and can be computed.  Calculating explicitly the first few terms in
the operators $\hln$ for small $|n|$, we have
\begin{eqnarray}
\hat{L}_3 & = & - \frac{\pdv}{\pdv z_3} + \cd_4, \nonumber \\
\hat{L}_2 & = & - \frac{\pdv}{\pdv z_2} - \frac{1}{2} z_1
\frac{\pdv}{\pdv z_3} + \cd_4, \nonumber \\
\hat{L}_1 & = & - \frac{\pdv}{\pdv z_1} + \frac{1}{2} z_2
\frac{\pdv}{\pdv z_3} + \cd_4, \nonumber \\
\hat{L}_0 & = & h + z_1 \frac{\pdv}{\pdv z_1} + z_2 \frac{\pdv}{\pdv
z_2} + z_3 \frac{\pdv}{\pdv z_3} + \cd_4,   \\
\hat{L}_{-1} & = & -2 h z_1 + (3z_2 - z_1^2) \frac{\pdv}{\pdv z_1} + (4
z_3 - 2 z_1 z_2) \frac{\pdv}{\pdv z_2} + \co(z_4), \nonumber \\
\hat{L}_{-2} & = & -(4h + \frac{c}{2}) z_2 + 3 h z_1^2 + (5 z_3 -
\frac{13}{2} z_1 z_2 + z_1^3) \frac{\pdv}{\pdv z_1} + \co(z_4), \nonumber \\
\hat{L}_{-3} & = & - (6h + 2c) z_3 + (13h + c) z_1 z_2 - 4h z_1^3 +
\co(z_4), \nonumber \end{eqnarray}
where $\cd_4$ denotes terms containing derivatives $\pdv / \pdv z_k$
with $k \geq 4$, and $\co(z_4)$ denotes terms containing
quasi-homogeneous polynomials
of degree at least 4.  From these formulae, we can explicitly compute
the lowest degree states arising from the action of the Virasoro algebra
on $\nul$.  Using a basis for $R$ of monomials in the variables $z_i$,
where $f \in R$ is represented by the state $|f\rangle$, we have
\begin{eqnarray}
\hat{L}_{-1} \nul & = & -2 h |z_1\rangle, \nonumber \\
\hat{L}_{-1}^2 \nul & = & -6h |z_2\rangle + (4 h^2 + 2h) |z_1^2\rangle,  \\
\hat{L}_{-2} \nul & = & -(4h + \frac{c}{2}) |z_2\rangle + 3 h
|z_1^2\rangle. \nonumber
\end{eqnarray}
Since $\nul$ is the unique highest weight state in $R$, we expect that
for those values of $h,c$ with a vanishing Kac determinant at level
$k$ (i.e.,
where the algebraic representation contains a null state at level
$k$), we should
find linear dependences among the states generated by combinations of
raising operators of degree $k$.  As an example, from the above
expressions for $\hat{L}_{-1}^2 \nul$ and $\hat{L}_{-2} \nul$, it is
easy to see that these states are linearly dependent when
\bge
h = \frac{5 - c \pm \sqrt{(c-1)(c-25)}}{16}.\label{eq:red} \ee
This is exactly the condition for the Kac determinant to vanish at
level 2.  We see then, that the structure of the representations
corresponding to the discrete series of representations with $c \leq
1$ is exactly that predicted by Aldaya and Navarro-Salas
in \cite{ANS1}.  In these cases, the carrier space $R$ of the
representation has only
one highest weight state, however it is not irreducible.  To achieve
an irreducible representation, one must take the orbit of $\nul$ in
$R$ under the action of the enveloping algebra (i.e., the space
spanned by the set of
states in $R$ of the form $(\prod \hat{L}_{-n_i}) \nul$.
Note that the reducibility of these representations occurs in a manner
quite different from that in which the usual algebraically constructed
representations are reducible.  In the algebraic construction, certain
linear combinations of states give null vectors, which generate Verma
modules which must be divided out to get an irreducible
representation.  In our case, as in that of Aldaya and Navarro-Salas,
there are no null states; the elements of the enveloping algebra which
give null states in the algebraic theory, give zero when applied to the
highest weight vector of our Virasoro module.  This property is
common to many Fock space representations of infinite-dimensional
algebras \cite{Thorn}.

We will now discuss briefly the possibility of realizing a unitary
structure on the coadjoint orbit representations we have constructed.  A
first step in describing such a structure would be to find an
expression for a Hermitian metric $\exp(H)$ on $\lbc$.  If we assume
that the connection we have defined by our gauge-fixing procedure is
the associated metric connection, as was the case for $SU(2)$,
then we can explicitly calculate $H$ as a formal power series in the
$z$'s.  We have
\begin{eqnarray}
A_0 & = & - f_0 + i \Phi_0 \\
& = & - h -i\langle (b_0, ic), \exp(- \sum_{n>0} \mu_nL_{-n})
\exp(- \sum_{n>0} z_n L_n) L_0 \exp(\sum_{n>0} z_n
L_n)\exp(\sum_{n>0} \mu_n L_{-n})  \rangle \nonumber \\
& = & \!\!\sum_{\sss k,l>0; \{n, m\}} \! \frac{(-1)^{k+l}}{k!\:l!} z_{n_1}
\ldots z_{n_k} \mu_{m_1} \ldots \mu_{m_l}
C_0(n_1,\ldots,n_k,-m_1,\ldots,-m_l)
(h+\frac{c}{24}(m_l^2-1)),\nonumber \end{eqnarray}
where the sum is taken over all $n_1, \ldots, n_k, m_1, \ldots, m_l >
0$ satisfying $n_1 + \ldots + n_k =  m_1+  \ldots + m_l$.
If we assume $A_{\bar{z}_n} = 0$ and $A_{z_n} = \pdv H / \pdv z_n$,
then we have
\bge
A_0 = - \sum_{n>0} n z_n A_{z_n},\ee
and
\bge
H = - \xi_0^{-1} A_0,\ee
up to a constant.  We will take this function $H$ as a candidate for the
Hermitian metric on $\lbc$.  The first few terms in a power series
expansion of $H$ are given by
\begin{eqnarray}
\lefteqn{H = -\sum_{n > 0} 2n(h + \frac{c}{24}(n^2-1)) |z_n|^2}
\nonumber \\
& & + \sum_{n,m>0}
\left[ (m^2 + 4mn + n^2) h + \frac{c}{24}(m^4 + 2m^3n + 2 m n^3 + n^4
- m^2 - 4 mn - n^2)\right] \nonumber \\
& & \;\;\;\;\;\;\;\;\;\;\;\;\;\times (\frac{z_m z_n \bar{z}_{m+n} + \bar{z}_m
\bar{z}_n z_{n+m}}{2}) + \co(z^4). \label{eq:hvir} \end{eqnarray}
Note that $H$ is expected to be real, in order to be a Hermitian
metric.

To have a complete description of a unitary structure on $R$, it would
now be necessary to find an invariant metric on $\di$.  Unfortunately,
it is unclear whether such a metric can be found.  Attempting to
describe such a metric as a formal power series gives rise to an
expression with divergent coefficients.  The matter is complicated by
the fact that the adjoint representation of the Virasoro algebra is not a
highest-weight representation.  It seems that some kind of
regularization scheme may be necessary to construct such a metric in a
sensible fashion.  We can, however, get some information about when
such a
unitary structure is likely to be possible directly from
(\ref{eq:hvir}).  If we take only the first term in  (\ref{eq:hvir}),
and approximate the metric with a Gaussian, we see that for $h \ll 0$ or
$c \ll 0$, the metric diverges badly, and we will certainly not find a
unitary structure.  When $h, c \gg 0$, a sensible
inner product on $R$ can be found, at least in perturbation theory, by
taking a product of Gaussian integrals.  Using the Hermitian metric
(\ref{eq:hvir}) to compute anything nonperturbative, however, would be
a difficult proposition.  Further progress in this direction will
be impossible until some sort of a regularized invariant metric on $\di$ can be
described explicitly.

Despite the fact that we cannot construct a unitary structure on $R$
by integrating a Hermitian metric on $\lbc$ over $\di$, we can still
use the more simple-minded approach mentioned in section 2 of
constructing a unitary structure on $R$ in the same fashion as is done
in the algebraic approach.  That is, let us assume that $\langle \: |
\: \rangle = 1$ and $\hat{L}_n^{\dag} = \hat{L}_{-n}$.  (We will use
physicists notation for the inner product on $R$, denoting the inner
product of two functions $f$ and $g$ by $\langle f | g \rangle$.)  We
can algebraically compute $\langle f | g \rangle$ whenever $f$ and $g$
are in the orbit of $\nul$ under the action of the $\hat{L}_{-n}$'s.
This calculation is equivalent to that usually carried out in the
algebraic construction (for a review and further references see
\cite{Gins}).  The standard result is that when $h \geq 0$ and $c \geq
1$, or when
\bge
c = 1 - \frac{6}{m(m+1)}, \label{eq:min1} \ee
and
\bge
h = \frac{[(m+1)p - mq]^2 - 1}{4 m (m+1)}, \label{eq:min2} \ee
for integers $m,p$ and $q$ satisfying $m > 2$ and $1 \leq q \leq p
\leq m-1$, the inner product
$\langle f | g \rangle$ is positive definite on the space defined by
the orbit of $\nul$ in $R$.  When $h \geq 0$ and $c \geq 1$, the
representation of the Virasoro algebra on $R$ is irreducible, so this
construction
gives a positive-definite inner product on $R$, with respect to which
$R$ is a Hilbert space.  For the representations where (\ref{eq:min1})
and (\ref{eq:min2}) are satisfied, the situation is slightly different.
In these cases this argument only tells us that the subspace $\ch$
given by the orbit of $\nul$  is a Hilbert space ($\ch$ carries the
irreducible highest
weight representation).  If we attempt to
extend the inner product from $\ch$ to $R$, while maintaining the
relationship $\hat{L}_n^{\dag} = \hat{L}_{-n}$, we get a contradiction.
For example, when $m=3, p=2, q=1$, and $h = c = 1/2$, Equation
\ref{eq:red} is satisfied, and we have a reducible representation on
$R$.  In this case we can compute
\bge
\langle z_1 | z_1 \rangle = \langle \: | \hl_1 \hl_{-1} | \: \rangle =
2h = 1 \ee
\[
\langle -\frac{9}{4} z_2 + \frac{3}{2} z_1^2 | -\frac{9}{4} z_2 +
\frac{3}{2} z_1^2 \rangle = \langle \: | \hl_2  \hl_{-2} | \: \rangle =
9/4. \]
If we attempt to extend this inner product to $R_2$, however, we get
\begin{eqnarray}
\langle 3z_2 - 2z_1^2 | z_1^2 \rangle & = & \langle z_1 | \hl_1  |
z_1^2 \rangle =
-2 \nonumber \\
& = & - \frac{4}{3}\langle \: | \hl_2  | z_1^2 \rangle = 0.
\end{eqnarray}
Thus, the inner product on $\ch$ cannot be extended to one on $R$.

In summary, we have found that for $h \geq 0, c \geq 1$, a Hilbert
space structure exists on $R$, with respect to which our
representations are unitary.  For the discrete series of
representations given by (\ref{eq:min1}) and (\ref{eq:min2}), the
subspace $\ch$ of $R$ given by the orbit in $R$ of $\nul$ carries an
irreducible unitary representation, however the Hilbert space
structure on $\ch$ does not extend to $R$.  For all other
representations, there cannot exist a Hilbert space structure on $R$,
as no other values of $h$ and $c$ allow unitary representations of
the Virasoro algebra.  Thus, we see that all unitary representations
of the Virasoro algebra can
be found in the class of representations we have constructed, although
we are unable to show that the unitary structure on these
representations can be described geometrically as arising from a
Hermitian metric on the line bundle $\lbc$.  This situation parallels
that of loop groups, where it is impossible to find an invariant
metric on the homogeneous space $LG/T$, but where a unitary structure
can be found for the representations on spaces of holomorphic
sections of line bundles over $LG/T$ in a manner similar to that we
have used here \cite{PS}.

\section{Conclusions}
\baselineskip 18.5pt

We have succeeded in constructing, for arbitrary $h$ and $c$, a
representation of the Virasoro algebra on the space $R$ of polynomials
in a countable set of variables $z_1, z_2, \ldots$.  The
Virasoro generators in these representations are given by
\begin{eqnarray}
\hln & = & \xn + f_n  \\
\hat{C} & = & c, \nonumber \end{eqnarray}
where the operators $\xn$ are first-order differential operators,
defined by (\ref{eq:xiform}), and the functions $f_n$ are
quasi-homogeneous polynomials in $R$ of degree $-n$, given by (\ref{eq:fdef}).
When $\frac{c - 24h}{c}$ is not the square of a nonzero integer, these
representations correspond to coadjoint
orbits of $\vir$ of the form $\di$.  For the exceptional values of $h$
and $c$, these representations are still defined, although they are
not coadjoint orbit representations.  The representations in the
case $h = c/24$ appear to be related to the natural Hilbert space for the
quantum field theory defined by Alekseev and Shatashvili in
\cite{al-sh1} which corresponds to Polyakov's theory of 2d gravity.

The results given here agree with previously known information about
the $\di$ Virasoro coadjoint orbits.  The characters of the
representations we have constructed are clearly given by
\bge
{\rm Tr}_{\raisebox{-.2ex}{\small R}} q^{\hat{L}_0} =
\sum_{n}( {\rm dim} \; R_n) q^{h + n} = q^h \prod_{n=1}^{\infty}
\frac{1}{(1-q^n)},\ee
since the carrier space of all these representations is $R$.  This
agrees with the result calculated by Witten in \cite{Witt1} using
perturbative and index theory techniques.  For those values of $h,c$
corresponding to the $c \leq 1$ discrete series, we find that $R$
contains only a single highest weight state, but that the
representation is reducible.  The irreducible representation is
achieved by taking the orbit of the highest weight state in $R$.  This
result agrees precisely with the predictions of Aldaya and
Navarro-Salas based on their group approach to quantization.

One interesting feature of the Virasoro representations we have
constructed is the fact that all the generators act as first-order
differential operators on the space $R$.  This feature distinguishes
the representations we have developed here from the well known
``Feigen-Fuchs'' free
field representations which were described in the work of
Dotsenko and Fateev \cite{DF} using a Coulomb
gas-like free field theory with background charge.  In the Feigen-Fuchs
representations, there is a set of operators $\{a_n: n \in \ZZ\}$,
corresponding to
free-field modes.  These operators satisfy the Heisenberg algebra
\bge
[ a_n, a_m] = 2 n \delta_{n, -m}.\ee
The operators $a_n$ act on a bosonic Fock space with a vacuum
$\nul_{\raisebox{-.2ex}{\small f}}$
satisfying $a_n \nul_{\raisebox{-.2ex}{\small f}} = 0$ for $n > 0$ and $a_0
\nul_{\raisebox{-.2ex}{\small f}} = 2 \alpha
\nul_{\raisebox{-.2ex}{\small f}}$.  The Virasoro generators appear as modes
of the stress-energy
tensor, and are written in terms of the $a$'s as \cite{Feld}
\begin{eqnarray}
L_n & = & \frac{1}{4} \sum_{k = -\infty}^{\infty} a_{n-k} a_k - \alpha_0
(n+1) a_n,\;\;\;{\rm for} \; n \neq 0,  \\
L_0 & = & \frac{1}{2} \sum_{k = 1}^{\infty} a_{-k} a_k + \frac{1}{4}
a_0^2 - \alpha_0 a_0. \nonumber \end{eqnarray}
These generators satisfy a Virasoro algebra with
\bge h = \alpha(\alpha - 2 \alpha_0),\;\;\;\; c = 1 - 24 \alpha_0^2.\ee
We can reinterpret the bosonic Fock space in terms of the space $R$ by
defining
\begin{eqnarray}
a_n & = & 2 n \frac{\pdv}{\pdv z_n}, \;\;\;{\rm for} \; n > 0,
\nonumber \\
a_0 & = & 2 \alpha, \\
a_{-n} & = & z_n, \;\;\;{\rm for} \; n > 0. \nonumber \end{eqnarray}
In this notation, the Virasoro generators are
\begin{eqnarray}
L_n & = & \sum_{k = 1}^{n-1} k (n-k) \frac{\pdv}{\pdv z_k} \frac{\pdv}{\pdv
z_{n-k}} + 2 n (\alpha - \alpha_0 (n+1)) \frac{\pdv}{\pdv z_n} +
\sum_{k = n+1}^{\infty} k
z_{k-n} \frac{\pdv}{\pdv z_k}, \;\;\; {\rm for} \; n > 0 \nonumber \\
L_0 & = & \alpha(\alpha - 2 \alpha_0) +
\sum_{k = 1}^{\infty} k
z_k \frac{\pdv}{\pdv z_k},  \\
L_{-n} & = & (\alpha + \alpha_0(n-1)) z_n + \frac{1}{4} \sum_{k =
1}^{n-1} z_k z_{n-k} + \sum_{k = 1}^{\infty} k
z_{k+n} \frac{\pdv}{\pdv z_k}, \;\;\; {\rm for} \; n > 0. \nonumber
\end{eqnarray}
For $n > 0$, these generators are second order differential
operators on $R$.  These representations seem to be
fundamentally different from those we have constructed by the coadjoint
orbit method.  Note that the Feigen-Fuchs
representations have null states (i.e., states $\nuls$ other than the
vacuum satisfying $L_n \nuls = 0$ for all $n > 0$)
for certain values of $\alpha$ and
$\alpha_0$.  Thus, the structure of these representations is in some
sense more complicated than that of the coadjoint orbit representations we
have described in this paper.  This fact also implies that it is not
possible in general to relate the Feigen-Fuchs representations to the
coadjoint orbit representations via a generalized Bogoliubov transformation.
Such a transformation would have to leave both the vacuum and the
grading of the Heisenberg algebra fixed, and would allow a null
state in the Feigen-Fuchs representation to be described as a highest
weight state in $R$, which cannot exist by proposition \ref{p:p4}.
The structure of the Fock space in the
Feigen-Fuchs representations was originally described in \cite{FF2}.
It was subsequently shown by Felder that the irreducible
representations in the $c \leq 1$ discrete series could be described
using a BRST-type
screening operator which was previously introduced by Thorn
\cite{Thorn}.  It is possible that a similar construction could be
realized for the coadjoint orbit representations we have described
here.

One issue we have not completely resolved in this paper is the question of
unitarity.  We have given a candidate for a Hermitian metric on the
line bundles associated with the coadjoint orbit representations, but
without an invariant metric on the space $\di$, it is impossible to
perform explicit calculations.
We have shown that an inner product can be defined on $R$ when $h \geq
0, c \geq 1$, or on a subspace $\ch \subset R$ when $h$ and $c$
correspond to the discrete series of unitary representations, with
respect to which our representations are unitary.  It is not clear,
however, whether this inner product can be related in any natural way
to the Hermitian metric on $\lbc$.  It is also unclear whether there
exists a natural (geometric) reason for the breakdown of unitarity at
$c=1$.  Hopefully the results presented here will motivate further
investigation of these questions and of the geometric approach to
conformal field theory in general.

\vspace{.3in}
{\Large{\bf Acknowledgements}}

The author would like to thank Orlando Alvarez for helpful discussions
throughout the progress of this work and for suggesting several
improvements in this manuscript,
and Korkut Bardak\c{c}i, Anton Kast, Nolan
Wallach, and Bruno Zumino for helpful conversations.

\end{document}